\documentclass[]{interact}

\usepackage{epstopdf}

\usepackage[authoryear,round]{natbib}

\usepackage{todonotes}
\usepackage[most]{tcolorbox}
\usepackage{paracol}
\usepackage{enumitem}
\usepackage{graphicx}
\usepackage{subcaption}
\usepackage{xcolor}
\usepackage{array}
\usepackage{colortbl}
\usepackage{booktabs}
\usepackage{framed}
\usepackage{placeins} 
\usepackage{xspace}
\definecolor{shadecolor}{gray}{0.9}

\theoremstyle{plain}

\usetikzlibrary{positioning,matrix}

\tcbset{
  userchat/.style={colback=cyan!8!white, colframe=cyan!60!black, fonttitle=\bfseries, title=User, arc=1mm, boxrule=0.7pt, left=2pt, right=2pt, top=2pt, bottom=2pt},
  assistantchat/.style={colback=orange!4!white, colframe=orange!60!black, fonttitle=\bfseries, title=Assistant, arc=1mm, boxrule=0.7pt, left=2pt, right=2pt, top=2pt, bottom=2pt}
}
\usetikzlibrary{positioning,calc,arrows.meta,backgrounds,fit,shapes.geometric}

\begin{document}

\title{A Benchmark to Assess Common Ground in Human-AI Collaboration}

\author{
\name{Christian Poelitz\textsuperscript{a}, Finale Doshi-Velez\textsuperscript{b} and Si\^{a}n Lindley\textsuperscript{a}}
\affil{\textsuperscript{a}Microsoft Research, Cambridge, UK; \textsuperscript{b}Harvard University, Cambridge, MA, USA}
}

\maketitle

\begin{abstract}
AI is becoming increasingly integrated into everyday life, both in professional work environments and in leisure and entertainment contexts. This integration requires AI to move beyond acting as an assistant for informational or transactional tasks toward a genuine collaborative partner. Effective collaboration, whether between humans or between humans and AI, depends on establishing and maintaining common ground: shared beliefs, assumptions, goals, and situational awareness that enable coordinated action and efficient repair of misunderstandings. While common ground is a central concept in human collaboration, it has received limited attention in studies of human–AI collaboration. In this paper, we introduce a new benchmark grounded in theories and empirical studies of human–human collaboration. The benchmark is based on a collaborative puzzle task that requires iterative interaction, joint action, referential coordination, and repair under varying conditions of situation awareness. We validate the benchmark through a confirmatory user study in which human participants collaborate with an AI to solve the task. The results show that the benchmark reproduces established theoretical and empirical findings from human–human collaboration, while also revealing clear divergences in human–AI interaction.
\end{abstract}


\newcommand{\TOOL}{\textsc{TONETREETOOL}\xspace}

\section{Introduction}
Common ground is the shared knowledge, beliefs, and assumptions between parties in a conversation, along with the awareness that this knowledge is shared. In human communication, establishing common ground has been shown to be essential for mutual understanding and coordinated action. Conversely, a lack of common ground can lead to misunderstandings, inefficiencies, and collaboration breakdowns.

Research has shown that AI models do not show the interaction patterns needed to build common ground in human-AI interaction, and that when models are adapted to perform more of these, human-AI collaboration improves~\citep{collabllm2025}. 

Drawing on this insight, we start with the proposition that building common ground will be essential if AI models are to support activities that go beyond providing assistance or performing well-defined transactions to people. Research has shown that negotiation of common ground can positively affect human team decision making when solving complex problems~\citep{Beers2006}, and that common ground is essential to joint problem solving. \citet{roschelle1995construction}~argue that when people work together to find novel solutions, they build a joint problem space, or an emergent, socially-negotiated set of knowledge elements, which includes goals, problem state descriptions, potential problem solving actions, and their relationships. This is predicated upon common ground. If AI is to serve as an effective counterpart in problem-solving, or indeed in any task that requires shared planning, coordination, or negotiation of meaning, it will need to be able to build common ground with people. This will become increasingly important as AI models are given more autonomy and agency, and act asynchronously and over longer time horizons. These changes, already underway, highlight the need for AI models to co-develop mutually acceptable plans, to understand, communicate, and adapt to changes, and to engage in the finding of novel solutions to problems that may emerge along the way. As common ground is dynamically co-constructed and maintained over time, effective collaboration depends on continuous mutual adaptation, and a shared history of interactions, rather than isolated, transactional exchanges. 

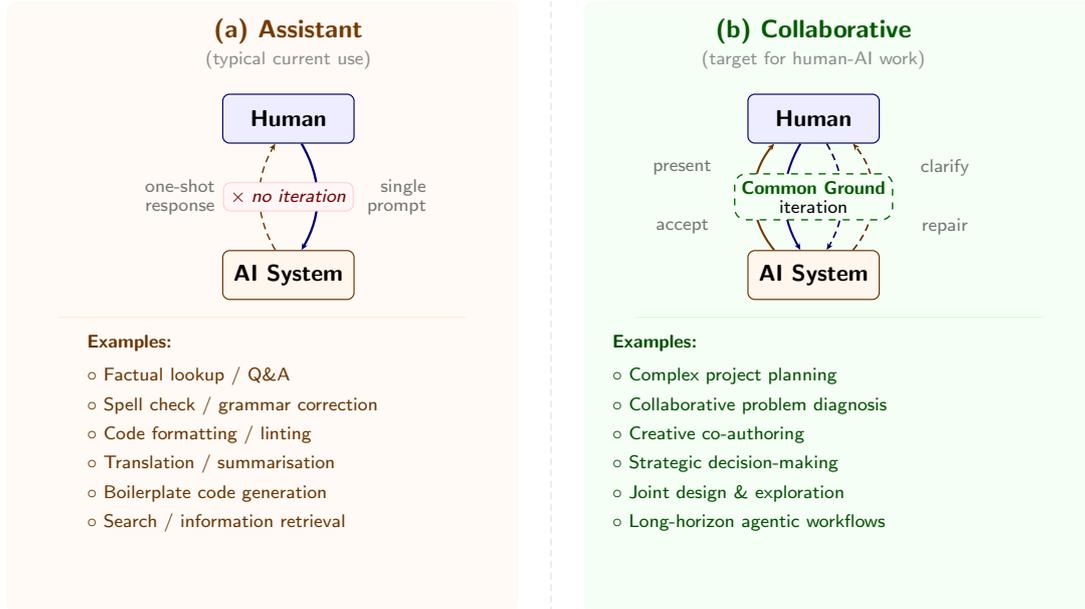
\begin{figure}[t]
\centering
\resizebox{\textwidth}{!}{%
\begin{tikzpicture}[
  font=\sffamily\small,
  >={Stealth[length=3pt,width=2.5pt]},
  actor/.style={draw, rounded corners=3pt, minimum width=2.0cm,
                minimum height=0.75cm, align=center,
                font=\sffamily\small\bfseries, line width=0.5pt},
  lbl/.style={font=\sffamily\scriptsize, text=black!55},
  item/.style={font=\sffamily\scriptsize, anchor=west},
]

\fill[orange!4, rounded corners=5pt] (-0.8, -3.8) rectangle (7.0, 5.6);

\node[font=\sffamily\normalsize\bfseries, text=orange!45!black]
  at (3.5, 5.15) {(a) Assistant};
\node[font=\sffamily\scriptsize, text=black!45] at (3.5, 4.7)
  {(typical current use)};

\node[actor, fill=blue!7, draw=blue!40!black]    (hA) at (3.5, 3.8) {Human};
\node[actor, fill=orange!7, draw=orange!40!black] (aA) at (3.5, 1.4) {AI System};

\draw[->, line width=1pt, blue!45!black]
  ([xshift=2mm]hA.south) to[out=-60, in=60] ([xshift=2mm]aA.north);
\node[lbl, align=right] at (5.15, 2.6) {single\\[-1pt]prompt};

\draw[->, line width=0.6pt, orange!40!black, densely dashed]
  ([xshift=-2mm]aA.north) to[out=120, in=-120] ([xshift=-2mm]hA.south);
\node[lbl, align=right] at (1.85, 2.6) {one-shot\\[-1pt]response};

\node[font=\sffamily\scriptsize, text=red!45!black,
      draw=red!20, rounded corners=3pt, fill=red!3,
      inner sep=3pt] at (3.5, 2.6)
  {\texttimes\;\textit{no iteration}};

\draw[orange!20, line width=0.4pt] (0.0, 0.75) -- (6.2, 0.75);

\node[font=\sffamily\scriptsize\bfseries, text=orange!40!black, anchor=west]
  at (0.3, 0.35) {Examples:};
\node[item, text=orange!40!black] at (0.3, -0.15) {$\circ$~Factual lookup / Q\&A};
\node[item, text=orange!40!black] at (0.3, -0.6)  {$\circ$~Spell check / grammar correction};
\node[item, text=orange!40!black] at (0.3, -1.05) {$\circ$~Code formatting / linting};
\node[item, text=orange!40!black] at (0.3, -1.5)  {$\circ$~Translation / summarisation};
\node[item, text=orange!40!black] at (0.3, -1.95) {$\circ$~Boilerplate code generation};
\node[item, text=orange!40!black] at (0.3, -2.4)  {$\circ$~Search / information retrieval};

\fill[green!4, rounded corners=5pt] (8.0, -3.8) rectangle (15.8, 5.6);

\node[font=\sffamily\normalsize\bfseries, text=green!35!black]
  at (11.5, 5.15) {(b) Collaborative};
\node[font=\sffamily\scriptsize, text=black!45] at (11.5, 4.7)
  {(target for human-AI work)};

\node[actor, fill=blue!7, draw=blue!40!black]    (hB) at (11.5, 3.8) {Human};
\node[actor, fill=orange!7, draw=orange!40!black] (aB) at (11.5, 1.4) {AI System};

\draw[->, line width=0.9pt, blue!45!black]
  ([xshift=-2mm]hB.south) to[out=-120, in=120] ([xshift=-2mm]aB.north);
\draw[->, line width=0.9pt, orange!45!black]
  ([xshift=-6mm]aB.north) to[out=130, in=-130] ([xshift=-6mm]hB.south);
\node[lbl] at (9.5, 3.05) {present};
\node[lbl] at (9.5, 2.15) {accept};

\draw[->, line width=0.7pt, blue!35!black, densely dashed]
  ([xshift=2mm]hB.south) to[out=-60, in=60] ([xshift=2mm]aB.north);
\draw[->, line width=0.7pt, orange!35!black, densely dashed]
  ([xshift=6mm]aB.north) to[out=50, in=-50] ([xshift=6mm]hB.south);
\node[lbl] at (13.5, 3.05) {clarify};
\node[lbl] at (13.5, 2.15) {repair};

\node[draw=green!40!black, line width=0.6pt, dashed,
      rounded corners=5pt, fill=green!4,
      inner sep=3pt, align=center,
      font=\sffamily\scriptsize]
  (cg) at (11.5, 2.6)
  {\textbf{\color{green!35!black}Common Ground}\\[-1pt]
   iteration};

\draw[green!20, line width=0.4pt] (8.8, 0.75) -- (15.0, 0.75);

\node[font=\sffamily\scriptsize\bfseries, text=green!30!black, anchor=west]
  at (8.3, 0.35) {Examples:};
\node[item, text=green!30!black] at (8.3, -0.15) {$\circ$~Complex project planning};
\node[item, text=green!30!black] at (8.3, -0.6)  {$\circ$~Collaborative problem diagnosis};
\node[item, text=green!30!black] at (8.3, -1.05) {$\circ$~Creative co-authoring};
\node[item, text=green!30!black] at (8.3, -1.5)  {$\circ$~Strategic decision-making};
\node[item, text=green!30!black] at (8.3, -1.95) {$\circ$~Joint design \& exploration};
\node[item, text=green!30!black] at (8.3, -2.4)  {$\circ$~Long-horizon agentic workflows};

\draw[black!12, line width=0.5pt, densely dashed]
  (7.5, -3.8) -- (7.5, 5.6);

\end{tikzpicture}%
}
\caption{%
Illustration of human–AI interaction tasks arranged by increasing need for common ground.
Left: Typical use cases of AI assistants, where the user specifies a task and the system produces a one-shot response. The input largely determines the desired output, ambiguity is minimal, and little or no mutual understanding is required; when misalignment occurs, the burden of repair falls primarily on the human.
Right: Use cases in which AI acts as a genuine collaborative partner. Human and AI engage in iterative dialogue, jointly presenting, clarifying, repairing, and accepting contributions while actively building and maintaining common ground. Such tasks require goals and scope to be negotiated, assumptions to be updated, and hypotheses to be jointly constructed and revised over time. Without mutual adaptation and shared understanding, a simple assistant cannot effectively support these interactions; misalignment accumulates and the human bears the full burden of repair~\citep{Clark1991GroundingIC}.}
\label{fig:intro}
\end{figure}

In this paper, we present a novel benchmark to assess the development of common ground in human-AI interaction. We provide initial results from the evaluation of one fixed AI model: GPT4.1. The benchmark centres on a joint task that requires iterative dialogues, actions, clarifications, and repairs in order to reach a solution, thereby potentially eliciting interaction patterns predicted by psychological theories of grounding and collaboration. The benchmark enables measurement of situation awareness, learning effects, communication efficiency, and grounding behaviours. Interaction patterns that have been previously identified as detrimental to grounding, including a tendency towards single-turn responses and overly long responses, asymmetric grounding effort that places the burden of alignment on the person, and shallow grounding behaviours that fail to produce mutually recognizable evidence of shared understanding~\citep{Clark1991GroundingIC, shaikh-etal-2024-grounding}, can be revealed through the task. Our aim is to not only explore surface-level grounding cues, but develop a method that helps elucidate whether common ground has genuinely been formed through a human-AI interaction.

Our main contributions are:
\begin{itemize}
    \item Review of prior work on common ground from social science theories of human communication and collaboration, that draws out patterns we might expect to see in human-AI interaction.
    \item Introduction of a new benchmark based on a problem-solving task, expected to underpin those interaction patterns.
    \item A confirmatory online study of human-AI interaction, to validate the new benchmark.
    \item Discussion of the results and their usefulness in measuring common ground in human-AI interaction.
\end{itemize}

\section{Background and related work}

\subsection{Human-AI collaboration}
Prior research that investigates human-AI interaction points to the potential importance of developing common ground. \citet{Dafoe2021} highlight the need for AI systems to develop social and collaborative capabilities, framing “Cooperative AI” as requiring understanding, communication, commitment, and shared norms. \citet{Bansal2019BeyondAT} emphasize the role of mental models in human–AI team performance, showing that decision-making can improve when people understand  AI’s reasoning. Further, \citet{andrews2023role} gives an overview of (shared) mental models and their importance in human-AI interaction, showing that, when their assumptions are incorrect, a person's mental model of the AI can influence the collaboration outcome~\citep{pataranutaporn2023influencing}. In addition, human's understanding of AI has been widely explored in the field of explainable AI~\citep{Mathew2025ExplainableAI}. Research shows that explanations can improve understanding, though the effect depends on factors such as the user’s expertise~\citep{10.1145/3397481.3450650} and mental load~\citep{10.1145/3630106.3659032} influencing how well the AI is understood. Complementing, researchers have investigated AI systems’ understanding of human mental states, often through of theory of mind experiments e.g., by \citet{ma2024holisticlandscapesituatedtheory, zhang2024mutualtheorymindhumanai}, testing AI's ability to infer the mental states of users. The results showed LLMs can struggle with situational understanding, especially conversationally, and often show understanding through actions. 

Importantly, failures to establish common ground - which must be jointly constructed and maintained by both humans and AI systems - appear to have measurable performance consequences. \citet{Vaccaro2024} conducted a systematic review and meta-analysis of existing experiments on human-AI systems to compare the performance of people, AI, and interactions between the two. They found that, on average, human–AI teams performed worse than the better-performing teams of either people or AI alone, particularly in decision-making tasks. Building on these results, \citet{berger2025fostering} argue that human–AI synergy depends on integrating outcome feedback, human learning, and iterative adaptation. Together, these findings highlight the need for shared understanding as part of human-AI interaction, with common ground potentially being a key aspect of this.
  
\subsection{Common ground}
Building common ground is essential to collective action between people and is increasingly being recognized as important to human-AI interaction. Common ground, as outlined by Clark and Brennan~\citep{Clark1991GroundingIC}, is proposed to be an interactive and collaborative process through which mutual understanding, for the purposes of a conversation, is established. Building common ground involves coordination of both content and process. Coordination of content builds upon mutual knowledge, mutual beliefs, and mutual assumptions. As gaps are identified, participants coordinate process to address misunderstandings and misalignments. The act of trying to establish that what has been said has been mutually understood and is therefore part of the common ground is called "grounding". 

Grounding involves the presentation and acceptance of information between conversation participants, to enable the continuous updating of common ground. People progress a conversation when the "grounding criterion" is met, i.e. when understanding is sufficient to meet the their current needs. Grounding cues can suggest problems, such as when people request clarification, or can demonstrate understanding, such as acknowledging what has been said or moving on. People ordinarily look for positive evidence of understanding, such as an acknowledgment, relevant next conversational turn, or continued attention. 

Clark and Brennan also suggest that grounding supports the principle of least collaborative effort; participants try to minimize their collaborative effort, or the work they do, to reach mutual understanding. Often it will take more collaborative effort for a speaker to produce a perfect message and for a listener to comprehend it as a single utterance, than for a speaker to present a provisional message and ask for confirmation, or for them to break a complex message into installments and check that understanding is gradually being built. Thus, grounding is a mutual behaviour that involves both signalling and checking. Through grounding, people engage in effective communication built on top of sufficient shared understanding to meet their current purpose. 

The process of grounding and its influence on the effectiveness of communication has been extensively studied in the social sciences. Early work by Krauss at al.~\citep{krauss1966concurrent, krauss1969development} showed how people built common ground in a task to describe novel design patterns over the telephone. In line with the principle of least collaborative effort, conversation partners developed linguistic anaphorical "references" to previously introduced pattern elements, enabling them to use fewer words as the conversation progressed. \citet{CLARK19861} emphasise that these references are collaboratively negotiated. Just as the speaker should check for understanding, the listener should signal understanding. So, if a reference is not understood, a listener should ask for clarification; if a speaker detects misunderstanding they should try to repair it. Once mutual understanding is reached, the reference is accepted as part of common ground. Both Krauss et al. and Clark at al. demonstrate that communication between pairs of people becomes more efficient over time, as shared referential language, which can be used as shorthand to refer to elements in a task, is developed. Efficiency is demonstrated via a) fewer words needed to describe task elements, and b) fewer acts of clarification and repair. Their research clearly demonstrates that pairs develop their own shared ways of referring to content, which, by becoming part of the common ground, increases communication efficiency. Research by \citet{fussell1992coordination} provides a further demonstration of how understanding is built between pairs when initial assumptions are misplaced. Participants often had partially incorrect beliefs about what each other knew, and relied on feedback for repair. \citet{fussell1992coordination} also show that participants often begin with minimalist descriptions and iteratively work towards information that is sufficient for understanding. However, while assumptions and references were developed through interaction, once grounded they tended to be stable and, as \citet{brennan1996conceptual} described, could remain so even if more efficient alternatives were available.

\subsection{Communication Medium and Common Ground}
How grounding is accomplished depends on the communication medium~\citep{Clark1991GroundingIC}. ~\citep{Clark1991GroundingIC} identify key factors that impact grounding: co-presence (physical presence), visibility (the ability to be seen), audibility (the ability to be heard), co-temporality (immediate turn-taking), simultaneity (simultaneous interaction), sequentiality (one turn following another), reviewability (the ability to review post-interaction), and revisability (the ability to edit). The costs of grounding, and therefore the degree of collaborative effort, are also shaped by the medium used. The medium can influence formulation, initial production, reception, understanding, setup for the entire interaction or collaboration, delays, speaker changes, asynchrony, display, faults, and repairs. For example, when the communication medium is speech, people can easily repeat back statements, or use backchannels (e.g., saying "uh-huh", or "yes"), to confirm understanding. When people are co-present, their joint actions and shared perceptual experiences support the building of common ground, which can function as shared bases for coordination and used as resources in conversation against which understanding can be checked \citet{clark1996using}. \citet{10.1145/358916.358947} showed that being collocated and seeing the immediate effects of actions in a remote bicycle repair task improved performance compared to remote collaboration using audio and video. However, having a video feed did not significantly improve performance over an audio-only channel, highlighting the need to consider what visual information to transmit in remote collaboration. A view of the task work area is often more valuable for establishing common ground that a view of someone's face. 

\citet{10.1145/587078.587084} studied further the effect of shared visual spaces on paired collaborative tasks. These helped maintain task awareness and supported grounding by reducing the effort needed to give feedback; visual feedback came for free. Reducing views of the shared visual space resulted in increased conversational grounding to compensate for missing visual feedback, demonstrated by more communication and longer task-completion times. \citep{Gergle01012013} extended these analyses, showing that visual information supports collaboration through two distinct mechanisms: conversational grounding and situation awareness. Again, visual information was found to support grounding and reducing this resulted in more confirmations, more detailed descriptions, and more requests for clarification, replicating \citet{10.1145/587078.587084} and supporting Clark’s concept of least collaborative effort. Additionally, the influence of situation awareness was found to be independent of conversational grounding. Visual information supported awareness of task state even when it failed to support grounding, for instance when visual perspectives were misaligned. This demonstrates that common ground requires shared understanding of visual information, not just access to it.

Text-based chat, while limiting some communication features that support grounding (such as backchannels~\citep{dideriksen2023quantifying} and provisional noun phrases~\citep{Clark1991GroundingIC}), does offer reviewability and revisability~\citep{Kraut2002Proximity}, allowing for the inspection of prior turns and more careful messaging, potentially reducing misunderstandings. As observed by \citet{Birnholtz2005GroundingNA}, if one participant in a conversation has more information than another, communicating that information in a carefully crafted message that can be reviewed repeatedly by the other party could be extremely beneficial to grounding. However, the principle of least collaborative effort suggests that people will only produce as much as text as they believe necessary. This requires some sort of alignment~\citep{koulouri2016and} on how to solve a task. Previous studies~\citep{sarkar-etal-2025-understanding} highlighted frictions in text-based goal-oriented dialogues between people, originating from divergent assumptions on how to solve a task coupled with insufficient signaling of misunderstandings and repairs. As noted earlier, grounding requires both coordination of content and coordination of process. The latter is likely to be especially relevant in complex, cooperative tasks~\citep{Convertino2008}.

The sequential structure of text-based interaction may also increase grounding costs by delaying error detection and correction, particularly when misunderstandings propagate across multiple turns. Research by \citet{monk2022common} and \citet{dillenbourg1996grounding} has explored combinations of text and shared workspaces, used to record potential solutions to the task in hand. Monk et al. measured recall of solutions and arguments for the solutions in solving a problem regarding the layout of a bank. While a shared report space improved recall of arguments, indicating better grounding, it had no significant effect on recall of solutions. They suggest that, when a shared workspace was not present, participants took on the extra costs of grounding solutions, these being their primary concern and subject to a strict grounding criterion of mutual acceptance. In contrast, participants accepted a lower criterion of negative evidence for arguments; unless there was a clear statement of lack of understanding or acceptance, these were reported. When the shared workspace was provided, it acted as a grounding constraint that reduced the lack of agreement concerning arguments. Dillenbourg et al. explore use of text in a MOO, where participants have different information needed to solve a problem. They note that grounding is influenced by factors including the probability that information is not shared, the cost of sharing it, and the cost if it is not shared. Participants perform grounding across modalities (e.g., moving from text chat to a shared whiteboard); if a grounding function cannot be performed in one mode, and really needs to be performed, it migrates to another. 

\subsection{Common Ground in human-AI interaction}
In the case of human-AI interaction, communication is predominantly via a chat environment comprising either written or spoken conversational turns. This may be contextualised by some resource, such as a document, webpage, or collection of artefacts. Typically, interactions in relation to these resources do not take the form of joint problem solving; AI models tend to play an assistive role, querying websites or collections, or generating content on command. Nevertheless, the content serves as a grounding constraint of sorts, and LLMs also perform some grounding acts through their utterances. For instance, recent studies by \citet{jones2026llms} and \citet{zeng2026lvlms} show that people and AI can converge on shared linguistic forms to refer to objects during the course of an interaction, and tools such as Copilot Researcher responds to an initial query by asking follow up clarification questions. Nevertheless, differences in how people and AI perform grounding are also evident; analyses by~\citep{shaikh-etal-2024-grounding, shaikh2025navigatingriftshumanllmgrounding} show a gap in how often clarifications are sought by people in comparison to AI models.

\citet{10.1108/INTR-06-2023-0514} identify social features and embodiment as mechanisms that are additional to grounding acts, mutual understanding, and shared mental models, when aiming to support the building of common ground in human–agent interaction. Likewise, \citet{Corti2016} showed that anthropomorphic design facilitates building common ground in human-AI interaction, as measured by an increase in willingness to engage in repairs. Here, people were less likely to repair misunderstandings when they believed they were working with AI rather than humans. These and other findings~\citep{tolzin2023designing} show that both the willingness of people and the design of AI systems are necessary if common ground is to be reached through human-AI interaction. 

However, it is important to acknowledge that there are also important differences between human interaction and human-AI interaction. It may not make sense to aim for mutual knowledge, mutual beliefs, and mutual assumptions in grounding between people and LLMs. AI models trained on human communication may perform grounding acts, for instance, by aligning on descriptions (\citet{jones2026llms}, \citet{zeng2026lvlms}) but it is not clear that these really signal the development of an underlying common ground. The performance of grounding acts can be achieved through language‑internal artifacts (e.g., repetition or lexical persistence) without requiring shared planning, coordinated action, or a shared goal. In the next section, we describe a benchmark that requires a person interacting with an AI partner to jointly act in a shared problem space, track evolving task state, repair errors, and coordinate progress over time. Grounding is evidenced through successful coordination of actions. This enables evaluation of common ground as a prerequisite for success, in addition to surface level grounding acts.

\section{Benchmark for common ground}
\label{sec:benchmark}
\begin{table}[htbp]  
\centering  
\small  
\begin{tabular}{p{2cm} p{3.2cm} p{3.5cm} p{4.2cm}}
\hline   
\textbf{Example study} &  
\textbf{Task} &  
\textbf{Key Mechanism} &  
\textbf{Main Findings} \\
\hline  

\multicolumn{4}{l}{\textit{Object level: descriptions, memory, perspective on tans, designs, photos, etc.}} \\  
\hline   
  
\citet{krauss1966concurrent} &  
Describing novel figures with feedback &  
Negotiation of shared references &  
Descriptions become shorter and more efficient over time  \\  
  
\citet{krauss1991perspective} &  
Describing abstract figures (social vs.\ nonsocial) &  
Perspective-taking and egocentric bias &  
Speakers overestimate shared knowledge; social descriptions are more accurate than nonsocial but still biased \\  
  
\citet{fussell1992coordination} &  
Referring to public figures in conversation &  
Probabilistic assumptions + online feedback &  
Speakers start with minimal descriptions and rely on repair \\  
  
\hline  
\multicolumn{4}{l}{\textit{Communication level: grounding processes and referential language to arrange objects}} \\  
\hline  
  
\citet{CLARK19861} &  
Tangram referential communication &  
Presentation--evaluation--acceptance cycles &  
References become shorter, definite, and require fewer turns over trials \\  
   
\citet{Clark1991GroundingIC} &  
Theoretical framework &  
Grounding criterion; least collaborative effort &  
Understanding must be established, not assumed  \\  
   
\citet{hh1981definite} &  
Theoretical analysis of definite reference &  
Copresence heuristics (physical, linguistic, indirect, community) &  
Definite reference appears effortless despite the theoretical requirement of infinite mutual knowledge \\ 
  
\hline  
\multicolumn{4}{l}{\textit{Task level: situation awareness and performance} on solving puzzle or repair tasks} \\  
\hline  
  
\citet{10.1145/587078.587084} &  
Bicycle repair task (collocated vs.\ remote) &  
Shared visual context &  
Collocated pairs completed tasks $\sim$25\% faster; video alone insufficient \\  
  
\citet{10.1145/587078.587084} &  
Collaborative puzzle solving &  
Shared visual workspace &  
Shared views reduce words, increase deixis, and improve performance \\  
  
\citet{Gergle01012013} &  
Collaborative puzzle solving &  
Situation awareness vs.\ conversational grounding &  
Grounding and awareness contribute independently \\  
  
\hline  
\end{tabular}  
\caption{Prior studies on common ground organized by level: references, communication, and task. While many studies look at several of the levels together, the table presents the studies associated with their main task and mechanisms of common ground. The main findings shortly describe the theoretical and experimental results described on the papers.}  
\label{tab:common-ground-related-work}  
\end{table}  

Current benchmarks for human-AI interaction mainly focus on simple transactional tasks, where interactions are limited to intent clarification or the provision of additional information needed to perform these transactions. 
In contrast, to measure the development of common ground, as defined by Clark, a benchmark will require a task that is genuinely collaborative, where partners aim to coordinate their actions to achieve a common goal. In this context, a collaborative task is one in which a person must work with an AI model to jointly and continuously construct, maintain, and repair assumptions and solutions through coordination (conversational grounding) and a shared understanding of the task state (situation awareness). Together, these processes enable the establishment and maintenance of common ground necessary for successful task completion.

Consequently, a benchmark should be designed around a genuinely collaborative task that is solved through human-AI interaction. To evaluate common ground as defined and studied in the foundational research described above (see Table~\ref{tab:common-ground-related-work} for a summary), such a benchmark must capture: (i) learning effects and efficiency gains at the task level, (ii) language adaptation at the communication level, and (iii) agreement on referential conventions and mutual understanding at the object (reference) level over time.

To operationalise these requirements, we design a matching task that follows the classic paradigm adopted in many studies of common ground in human communication. The task requires a pair to share information so that one member can recreate a pattern that the other has access to. The pieces of the pattern admit multiple plausible descriptions, enabling the emergence of shared referring conventions~\citep{krauss1966concurrent, CLARK19861}. The pieces are intentionally ambiguous, so as to enable cycles of presentation, repair, and acceptance~\citep{CLARK19861}. The task involves arranging these pieces into structured configurations, thereby requiring coordination around the referenced object~\citep{10.1145/587078.587084, Gergle01012013}. To measure learning effects and efficiency gains over time, the task involves multiple trials over which effects can emerge~\citep{CLARK19861, Gergle01012013}. Finally, we manipulate conditions of task awareness to examine trade-offs between shared visual context and conversational grounding~\citep{10.1145/587078.587084, Gergle01012013}. While prior research has used multiple variations of matching tasks, we most closely follow the puzzle task used by \citet{Gergle01012013}, as it also includes a comparison of two levels of task-state awareness.

\section{Benchmark design}
\begin{figure}[t]
\centering
\begin{minipage}{0.495\textwidth}
\centering
\includegraphics[width=\textwidth]{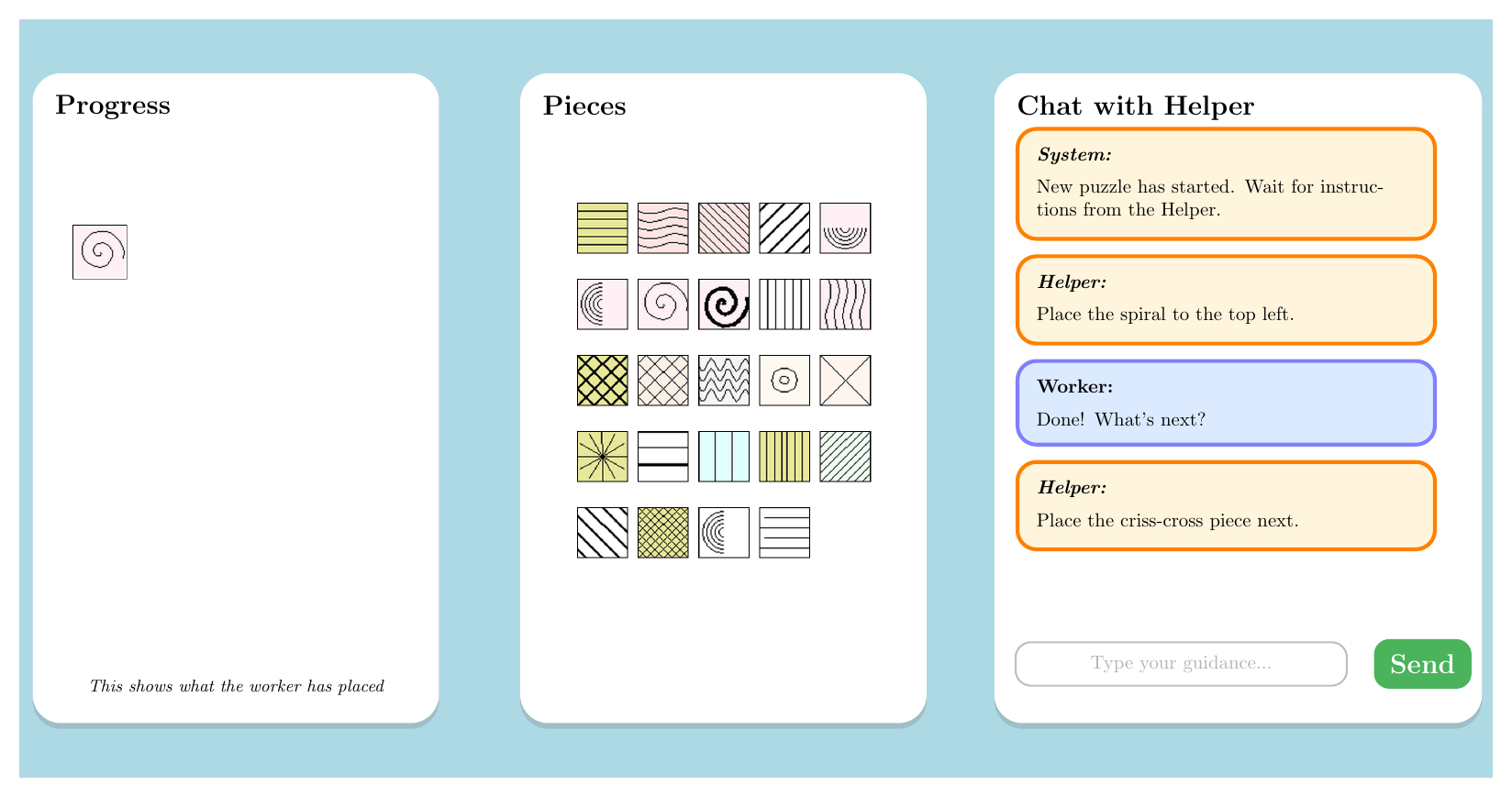}
\end{minipage}
\hfill
\begin{minipage}{0.495\textwidth}
\centering
\includegraphics[width=\textwidth]{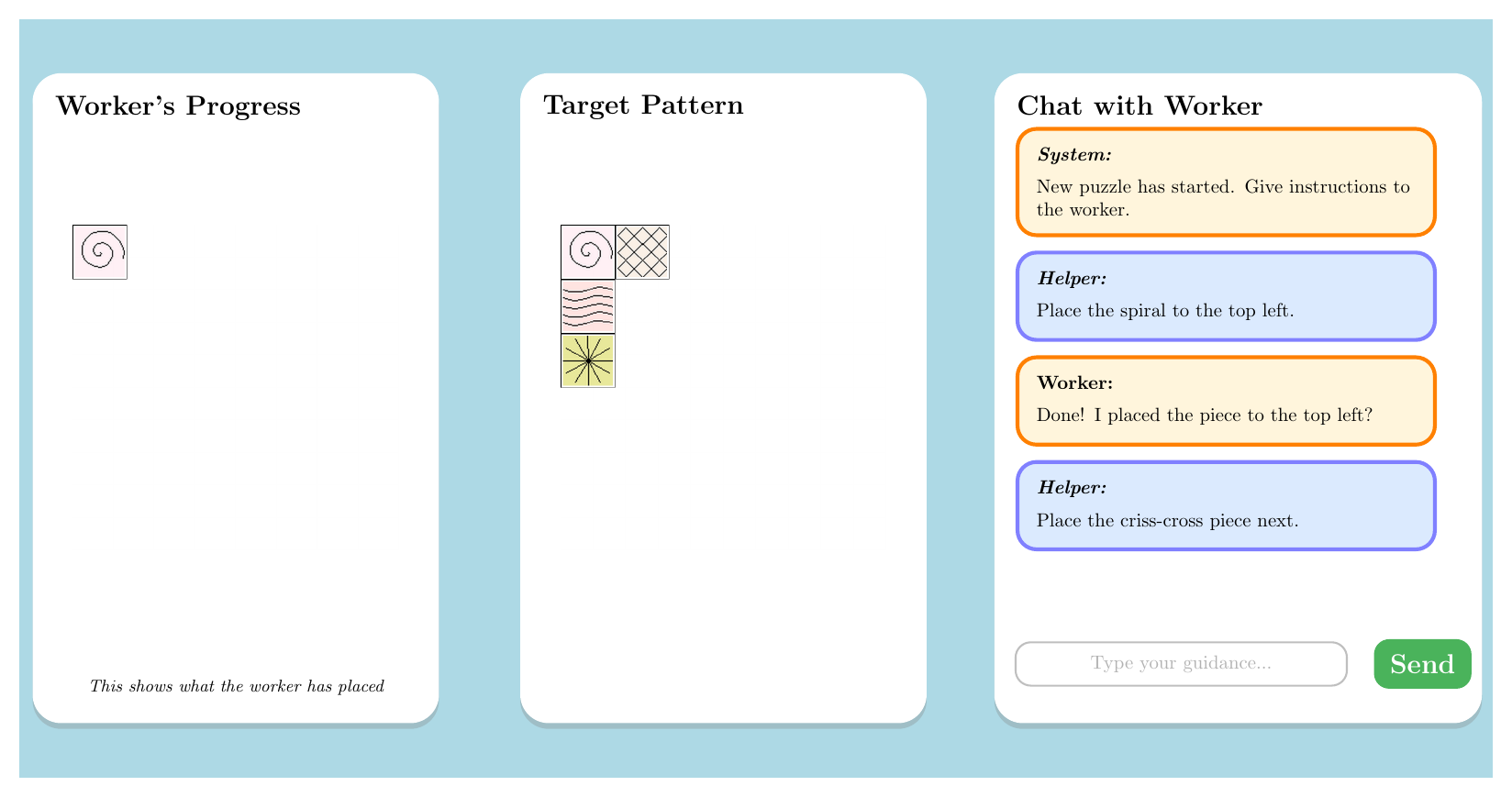}
\end{minipage}
\caption{Design of the puzzle task. Left: the Worker's view on solving the current puzzle with working area showing the current progress, available puzzle pieces and chat window. Right: the Helper's view with current Worker's progress (in the shared-view condition), the target puzzle and chat window. In the non-shared view condition, the current Worker's progress is removed from the Helper's view.}
\label{fig:task_views}
\end{figure}

\begin{figure}[t]  
  \centering  
  \begin{subfigure}[t]{0.23\textwidth}  
    \centering  
    \fbox{\includegraphics[width=\linewidth]{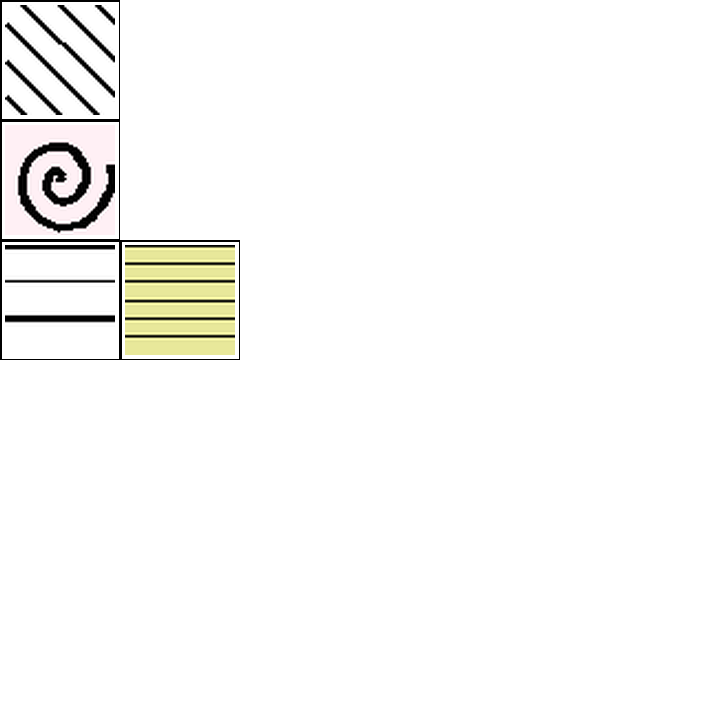}}  
    \caption{Target puzzle solution in trial 1}  
    \label{fig:trial1}  
  \end{subfigure}\hfill  
  \begin{subfigure}[t]{0.23\textwidth}  
    \centering  
    \fbox{\includegraphics[width=\linewidth]{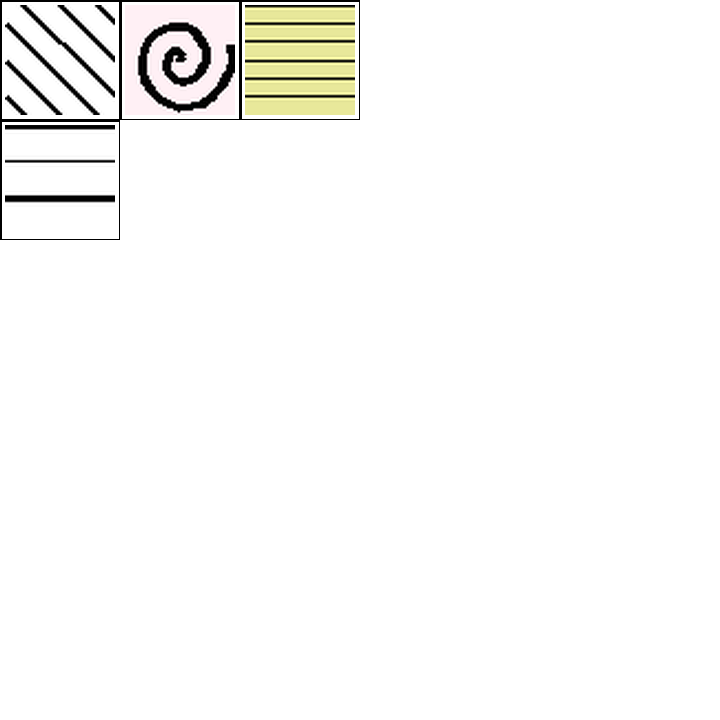}}  
    \caption{Target puzzle solution in trial 2}  
    \label{fig:trial2}  
  \end{subfigure}\hfill  
  \begin{subfigure}[t]{0.23\textwidth}  
    \centering  
    \fbox{\includegraphics[width=\linewidth]{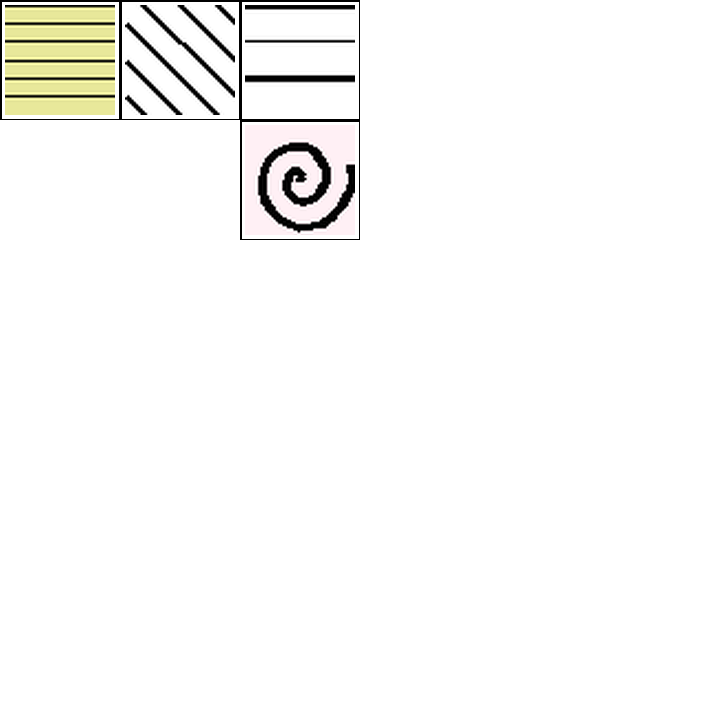}}  
    \caption{Target puzzle solution in trial 3}  
    \label{fig:trial3}  
  \end{subfigure}\hfill  
  \begin{subfigure}[t]{0.23\textwidth}  
    \centering  
    \fbox{\includegraphics[width=\linewidth]{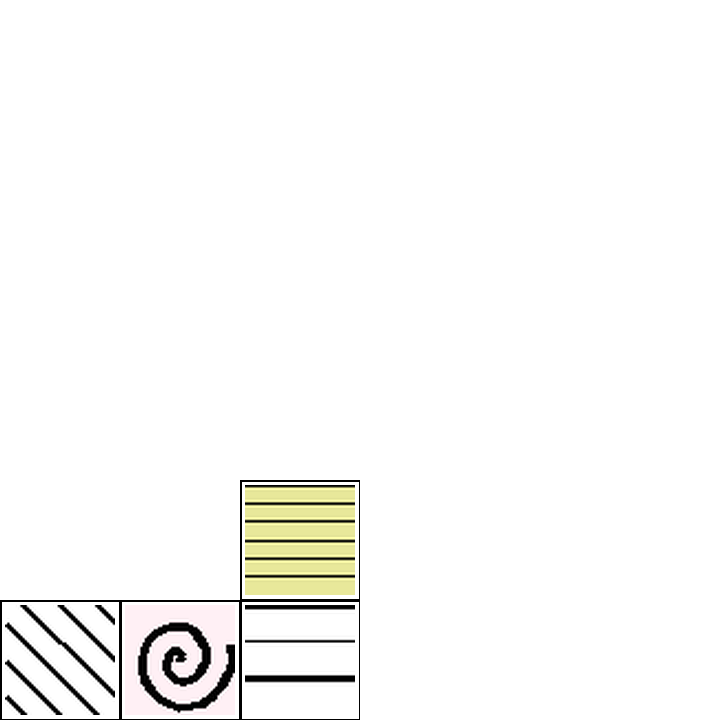}}  
    \caption{Target puzzle solution in trial 4}  
    \label{fig:trial4}  
  \end{subfigure}  
  
  \caption{Target puzzle solutions across trials 1-4 (a–d). The initial practice puzzle is shown in Fig.~\ref{fig:task_views}.}  
  \label{fig:trial_puzzles}  
\end{figure}  

Motivated by previous studies of common ground in human communication, and considering the requirements described in the last section, we now describe the design of our benchmark. The benchmark defines what is measured and how it is measured with the aim of supporting scientific research on common ground in human-AI interaction. 

The puzzle task is based on work by Gergle et al.~\citep{Gergle01012013} and requires both members of a pair to maintain awareness of the state of task objects and one another’s activities. In the task (see Fig.~\ref{fig:task_views}), a Helper (analogue to the 'Director' in \citet{CLARK19861}) instructs a Worker (analogue to the Matcher) on how to create a pattern consisting of four blocks. The Helper has access to a target pattern. The Worker has access to 24 pieces, many of which some have similarities to each other. The Helper does not have visibility of the full set of puzzle pieces in their view; the Worker did not have visibility of the target solution. To complete the puzzle, the Helper and Worker need to resolve the ambiguity around which pieces form part of the solution and where they should be placed. Puzzles are solved in series of four trials (see Fig.~\ref{fig:trial_puzzles} for the target puzzle solutions for each trial). The pieces that form part of the solution repeat over the four trials, although the pattern changes. Building shared understanding of this makes the interaction more efficient, enables the development of shared referential language over time, and improves task performance. 

Also following Gergle et al., we compare how making a view of the Worker's work area available to the Helper impacts interaction. Making the Worker's solution visible means that the Helper can take their actions, as well as (or instead of) their language, as evidence of understanding. When there is no shared view, the parties must completely rely on language to agree on how to progress and to determine whether a solution has been reached. 

To understand the impact of role on the joint construction and maintenance of common ground in human-AI interaction, some tasks are attempted with the AI as Worker and some are attempted with the AI as Helper. This symmetry ensures that the benchmark evaluates grounding behaviors and actions across complementary roles, rather than restricting the AI to an assistive role as in much prior work.

\section{Study - Benchmark validation}
\label{sec:validation}
In this section, we outline the setup of a confirmatory study of the development of common ground in human-AI interaction. In the study, one fixed LLM plays the role of either Helper or Worker when interacting with a person to solve the puzzle task described above. The purpose is to demonstrate that the benchmark can capture theoretically grounded aspects of common ground when they occur. We use GPT4.1 in our study since it is multi-modal (accepts both textual and visual inputs), but anticipate that future AI models will show different behaviours. A detailed description of the setup and the used prompts is reported in the Appendix ().

The confirmatory study is designed to investigate the following assumption using our benchmark:
\begin{shaded}
\noindent
We assume that for people to work with with AI to accomplish tasks where knowledge and/or action is asymmetrically distributed, they will need to develop and maintain common ground through human-AI interaction. 
\end{shaded}

The successful development of common ground should be reflected in similar patterns of human-AI interaction as seen in prior research of human communication~\citep{Clark1991GroundingIC, 10.1145/587078.587084, Gergle01012013}. In particular, on task performance, we expect to observe outcomes, not limited, but similar to:

\begin{itemize}
    \item When a shared view is present, actions will reveal misunderstandings and errors will be repaired more easily than when there is no shared view.
    \item When there is no shared view, pairs will use language more heavily than in the shared view condition, in an attempt to resolve ambiguity and successfully complete the puzzle. 
    \item However, some ambiguities will be left unresolved when there is no shared view. Pairs will make more errors in this condition than when they share a view. 
    \item Within each condition, a learning effect will occur resulting in fewer errors over trials.
    \item The amount of communication will reduce over trials in accordance with the principle of least collaborative effort.
\end{itemize}

More specifically, our benchmark should reveal whether people and AI models develop shared referential conventions~\citep{brennan1996conceptual} over time, and whether they can develop a shared understanding of the current state of a problem space. Prior work shows that people typically engage in these processes when building common ground (see above). We expect that the person will attempt to build shared references with the AI, and increasingly use definite references as shared understanding is build. We further expect the AI to surface similar behaviour, and that this will improve conversational efficiency and performance over time.

In this context, we further expect to observe the development of referential conventions as follows:
\begin{itemize}
    \item Pairs will develop shared language to refer to key puzzle pieces, reducing the number of characters per turn over trials (e.g., "place the spiral at the top left").
    \item Pairs will keep track of the state of the puzzle and use this to underpin references (e.g., "place the next piece to the right of the spiral").
    \item Pairs will build a shared understanding of how to solve the puzzle, reducing errors and the number of turns needed over a series of puzzles. E.g., pairs will understand that only 4 of the 24 pieces are relevant over subsequent turns.
\end{itemize}

Also, the benchmark should reveal the development of referential conventions as a process in which candidate references are presented, evaluated, clarified or repaired, and ultimately accepted~\citep{CLARK19861}. Therefore, we expect to observe the patterns of conversational grounding similar to:
\begin{itemize}
    \item Pairs will use clarifications and repairs to correct misunderstanding of presented references.
    \item Repairs and the need for clarifications will reduce over time as mutual understanding is build.
    \item When there is no shared view, pairs will increase conversational grounding resulting in more clarifications and repairs.
\end{itemize}

\subsection{Design}
We used a 2×2 between-subjects experiment with two conditions: Role (Helper vs. Worker) and View (shared view vs. non‑shared view). Participants interacted with the AI either as Helper or as Worker. The Helper either did or did not have access to a view of the Worker's work area, impacting whether or not they could use the Worker’s actions as evidence of (mis)understanding. 

\subsection{Participants}
\begin{figure}[t]
\centering
\begin{minipage}{0.45\textwidth}
\centering
\includegraphics[width=\textwidth]{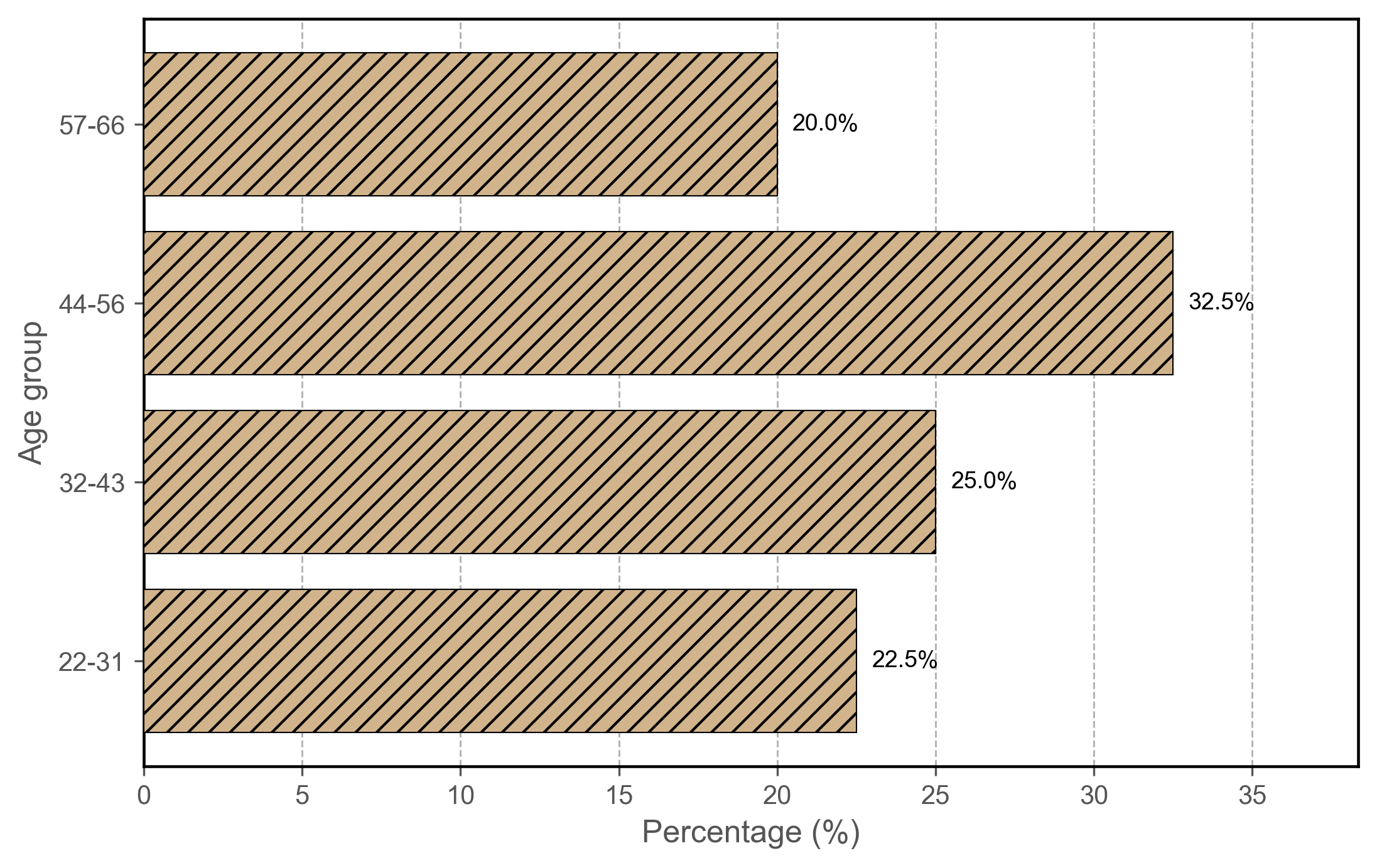}
 \caption*{(a) The age distribution of participants.}
\end{minipage}
\hfill
\begin{minipage}{0.45\textwidth}
\centering
\includegraphics[width=\textwidth]{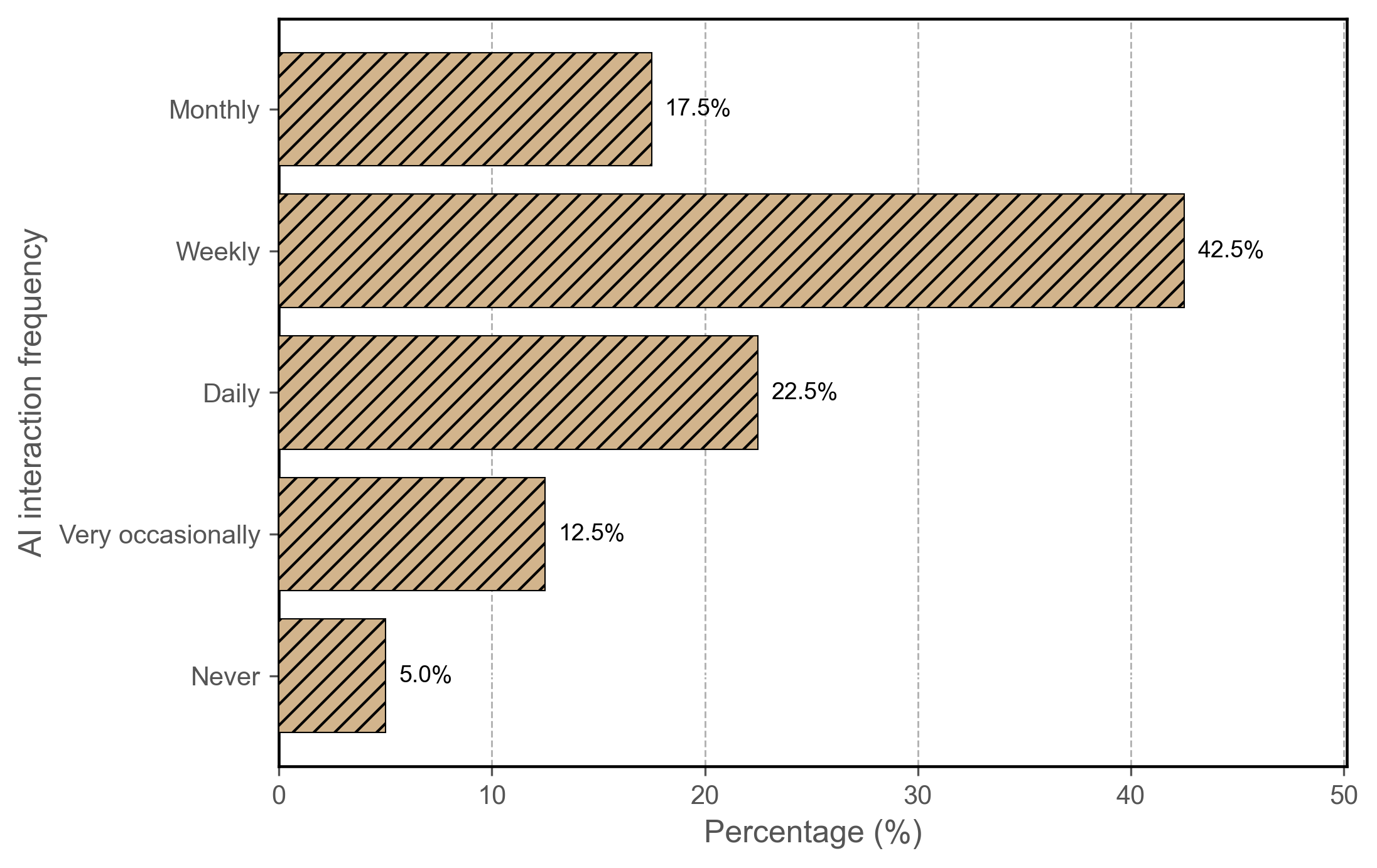}
\caption*{(b) The frequency of usage of AI.}
\end{minipage}
\caption{ Demographic distribution of participants by age and experience with AI. (a) The age distribution of participants. (b) The frequency of usage of AI.}
\label{fig:demographics}
\end{figure}

We recruited 
40 participants (17 male, 23 female)  via Prolific (www.prolific.com). There were 20 participants in each condition, with 10 participants playing each role. Participants were compensated directly via Prolific in line with platform norms (£12.5 per hour).

Participants were between 18 to 66 years of age, fluent in English, living in the UK, not diagnosed with any language-related disorders, not diagnosed with any visual impairments, and not diagnosed with any cognitive impairments. There were 17 male and 23 female participants. (See Fig.~\ref{fig:demographics} for the distributions of age and AI usage across participants based on a questionnaire after completing the task.)

\subsection{Materials}
Participants read a description of the study and consent form when signing up for the study. They read a set of instructions on how to complete the task before they began, and completed a questionnaire about their experience and demographics immediately afterwards. The task description contained all necessary information to solve the puzzles and included: the participant's role (Helper or Worker); the View condition they were in, (shared view or non-shared view); the number of puzzles they had to solve and how long the task was supposed to take (max. 5 minutes per puzzle). Participants also completed a questionnaire about their experience at the end of the task. 

The puzzle task was as described above: the Helper needed to instruct the Worker to place 4 puzzle pieces from a set of 24 in a target pattern. The Helper did not have visibility of the full set of puzzle pieces in their view; the Worker did not have visibility of the target solution. The task was completed via text-based interaction. In the shared view condition, a snapshot of the current solution was also sent to the Helper when the Worker indicated they had made a move by sending a message. 

We used OpenAI models hosted on Azure (GPT4.1 with vision capability). The participants used a simple web application like ChatGPT to interact with the model, with additional visual interfaces to either arrange the puzzle pieces as Worker or to see the target puzzle solution as Helper (see above). The web application was deployed on Azure. We logged all interactions (with the consent of the participants).

\subsection{Procedure}
Participants completed the consent form online ahead of the session. They were told that participation should not take longer than half an hour. The study involved completing a practice puzzle, followed by four puzzle trials (each block having repeating pieces to allow the development of common ground). Participants in the shared view condition had a snapshot of the work area  sent to the Helper whenever the Worker indicated they were ready to progress in the task; participants in the non-shared view condition had to rely solely on language to communicate. Pairs had to agree when they had solved the puzzle and so finished each trial. Once the pairs agreed they had completed the final puzzle, the human participant finished the task. 

\section{Data Analysis}
\label{sec:data_analysis}

\begin{table}[htbp]  
\centering  
\begin{tabular}{llll}  
\toprule  
\textbf{Statistic} & \textbf{Type} & \textbf{Instrument} \\  
\midrule  
Puzzle Success & Binary & Exact match with target solution \\  
Words per Trial & Numeric & Word count per turn \\  
Number of Turns & Numeric & Turn segmentation \\  
Noun Phrases & Multinomial & NLP-tool-based noun phrase extraction \\  
Referential Type & Multinomial & Regex-based classification \\  
Dialogue Acts & Multinomial & LLM-based annotation \\  
\bottomrule  
\end{tabular}  
\caption{Summary of analyzed variables, measurement types, and instruments.}  
\label{tab:analysis_summary}  
\end{table}  

The interactions between participants and the AI model as they attempted to solve the puzzle were logged. Logs included: (i) the dialogue messages between the participants and the AI, (ii) the actions performed by the Workers (person or AI), (iii) role assignments (Helper or Worker), and (iv) experimental condition (shared view or non-shared view). The data was grouped by condition, role, and trial (including practice trial), enabling analysis at the task level, communication level and object level. 

Our study is text-based and task completions time could be influence by LLM calls. Therefore, instead of measuring task completion time (similar to e.g., \citet{Gergle01012013}), we measure number of messages (turns) and message length (words) to assess communication effort and performance.

We used this data to compute simple statistics - such as word counts, turn counts, and exact puzzle matches - and linguistic measures, including noun phrases and dialogue acts, to asses how much the study results conform with the above described expected observations. The simple statistics are sufficient to test for the task-level expected observations about successful puzzle completions and communication effort (overall or across trials). To analyse language-specific expected observations - at communication and object levels - we use existing NLP tools and leverage LLMs as labeler for discourse markers~\citep{10.1162/tacl_a_00420}. 

Across all analyses, statistical significance was evaluated at a threshold of $0.05$. When applicable, we report test statistics and the corresponding p-values. Table~\ref{tab:analysis_summary} summarises the statistics we used, their measurement type, measurement instructions, and level for common ground (see Section~\ref{sec:benchmark}). 

In detail, we applied the following evaluation protocol to compare the outcome of the user study with the expected observations (see Section~\ref{sec:validation}):
\paragraph{Comparisons to expected observations on task performance (Task-level)}
For each trial, task success was measured as exact match between the final puzzle configuration and the target solution. To account for repeated measures and the binary type of the outcome, we use Generalized Estimating Equations (GEEs) with binominal family and logit link function. The outcome variable was conditioned on the condition (shared vs. non-shared view) and trial number, allowing us to test for differences between conditions and within condition, over trials (comparing the first puzzle outcome to all other outcomes). Significant results are reported as $\chi^2$ statistics and p-values.

\paragraph{Comparisons to expected observations on communication effort (Task-level)}
Communication effort was measured by the number of words used to solve the puzzle in each trial.  We used Mann-Whitney U tests to compare the average communications efforts between conditions. To measure learning effects over time, we fit linear mixed-effects models conditioned on condition and trial number, to account for random effects and the numeric types of the outcome. We use this to test whether there is a statistical reduction in communication effort over time indicating a learning effect. Significant results were reported as U-statistics (for the comparisons between conditions), F-statistics (for comparisons within condition between trials i.e., linear trends) and p-values.

\paragraph{Comparisons to expected observations referencing conventions (Object-level)}
To analyse referential conventions and their establishment over trials, we extracted noun phrases from the dialogue messages using an NLP tool (NLTK - Natural Language Toolkit) which parses the sentences in the messages. We then applied regular-expression based rules to filter for noun phrases which are referencing puzzle pieces only. The extracted noun phrases were aggregated across trials and categorized as (i) privately used by the human participant, (ii) privately used by the AI, or (iii) jointly used by both whilst solving the puzzle. This enables testing whether human and AI start using a shared vocabulary to refer to the puzzle pieces.

Each reference was further classified according to:
\begin{itemize}
    \item Definiteness: definite (e.g., the yellow piece) vs. indefinite (e.g., a yellow piece)
    \item Reference type: descriptive (e.g., the piece with stripes) vs. identifiers (e.g., piece 3 to location (1,2))
\end{itemize}

Changes in the distribution of reference types over trials are used as indicators of shared referential conventions, increasing communicative efficiency and mutual understanding. We used linear mixed-effects models to compare the distributions of noun phrases in the vocabularies and referencing styles across conditions and between trials.

\paragraph{Comparisons to expected observations on conversational grounding (Communication-level)}
To analyse conversational grounding, we follow \citet{CLARK19861} and locate presentation-acceptance cycles of object references. As noted by \citet{traum1998clark}, it is challenging to identify the boundaries of the presentation-acceptances cycles (e.g., acceptances can be recursive, creating new presentations). Consequently, instead of segmenting full cycles, each message was independently annotated for its grounding function~\citep{traum1992speech}.

Specifically, each turn was labelled as one of the following dialogue acts:
\begin{itemize}
    \item Presentation: First, a reference (i.e., the next puzzle piece) is presented in the chat e.g., "place the red piece at the top".
    \item Clarification: If the receiver identifies ambiguity they try to resolve this by asking for clarification e.g., "which red piece?".
    \item Repair: If either partner identifies (or recognises after clarification) potential misunderstanding or misalignment they try to fix this e.g., "the one with three stipes".
    \item Acceptance: Finally, if both partners recognise the reference as correct, it is accepted and can become part of the mutual understanding e.g., "Done. What is next?", "Now place the green piece [...]".
\end{itemize}
The messages were labelled automatically using a large language model (GPT5.2), following recent works demonstrating the reliability of LLMs for dialogue-act annotation \citep{10.1162/tacl_a_00420, shaikh-etal-2024-grounding, sarkar-etal-2025-understanding}. 

Changes in the distributions of grounding acts across trials and condition were used to assess whether pairs coordinated their communication and repaired misunderstandings more efficiently over trials. We used linear mixed-effects models to compare the distributions of grounding acts across conditions and between trials.

\section{Results}
In this section, we report the results from the user study to validate our benchmark as explained above. First, we analyse the results at the task-level.

\paragraph{Comparisons to expected observations on task performance (Task-level)}
\begin{figure}[t]
\centering
\begin{minipage}{0.48\textwidth}
\centering
\includegraphics[width=\textwidth]{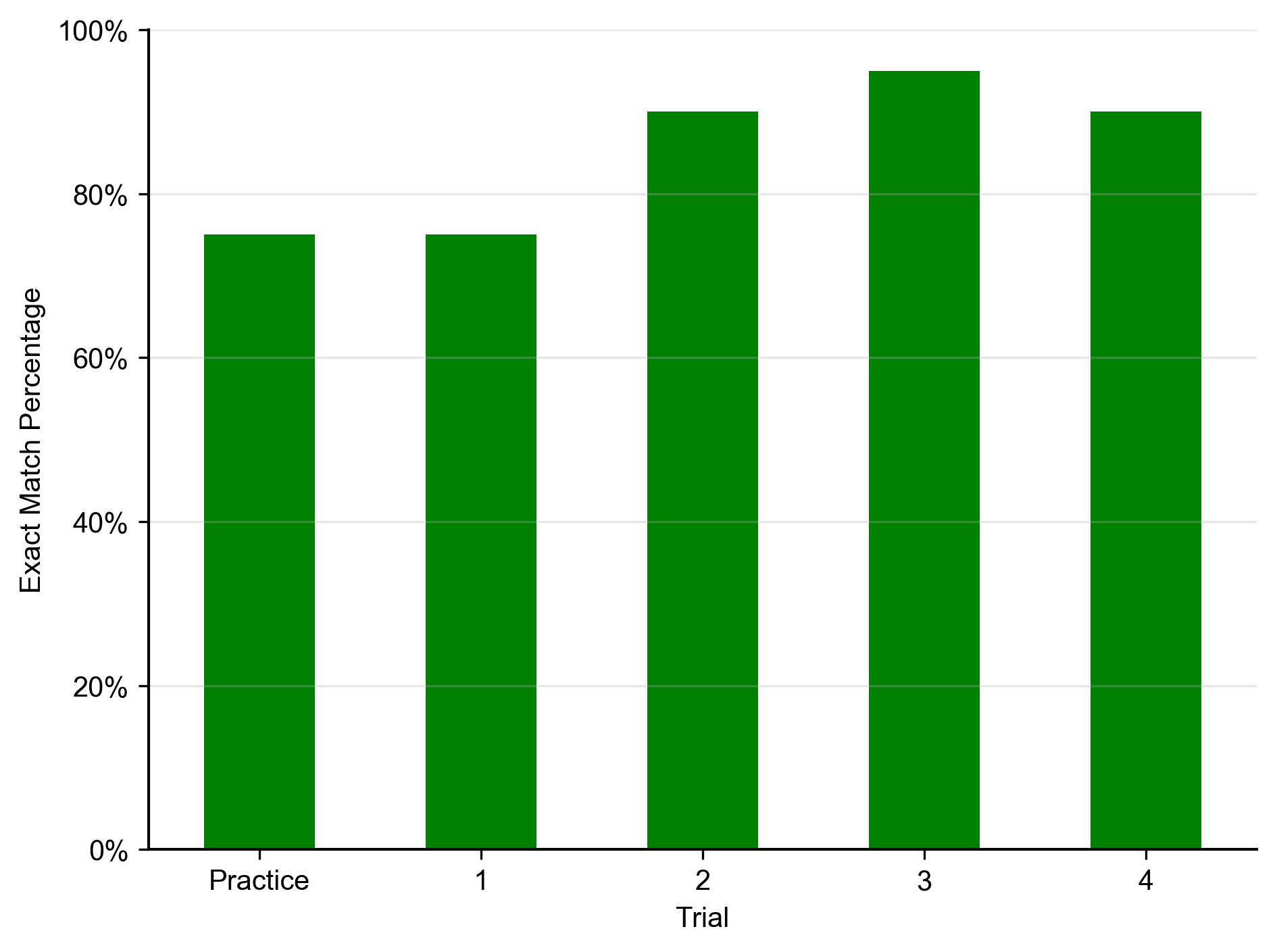}
\end{minipage}
\hfill
\begin{minipage}{0.48\textwidth}
\centering
\includegraphics[width=\textwidth]{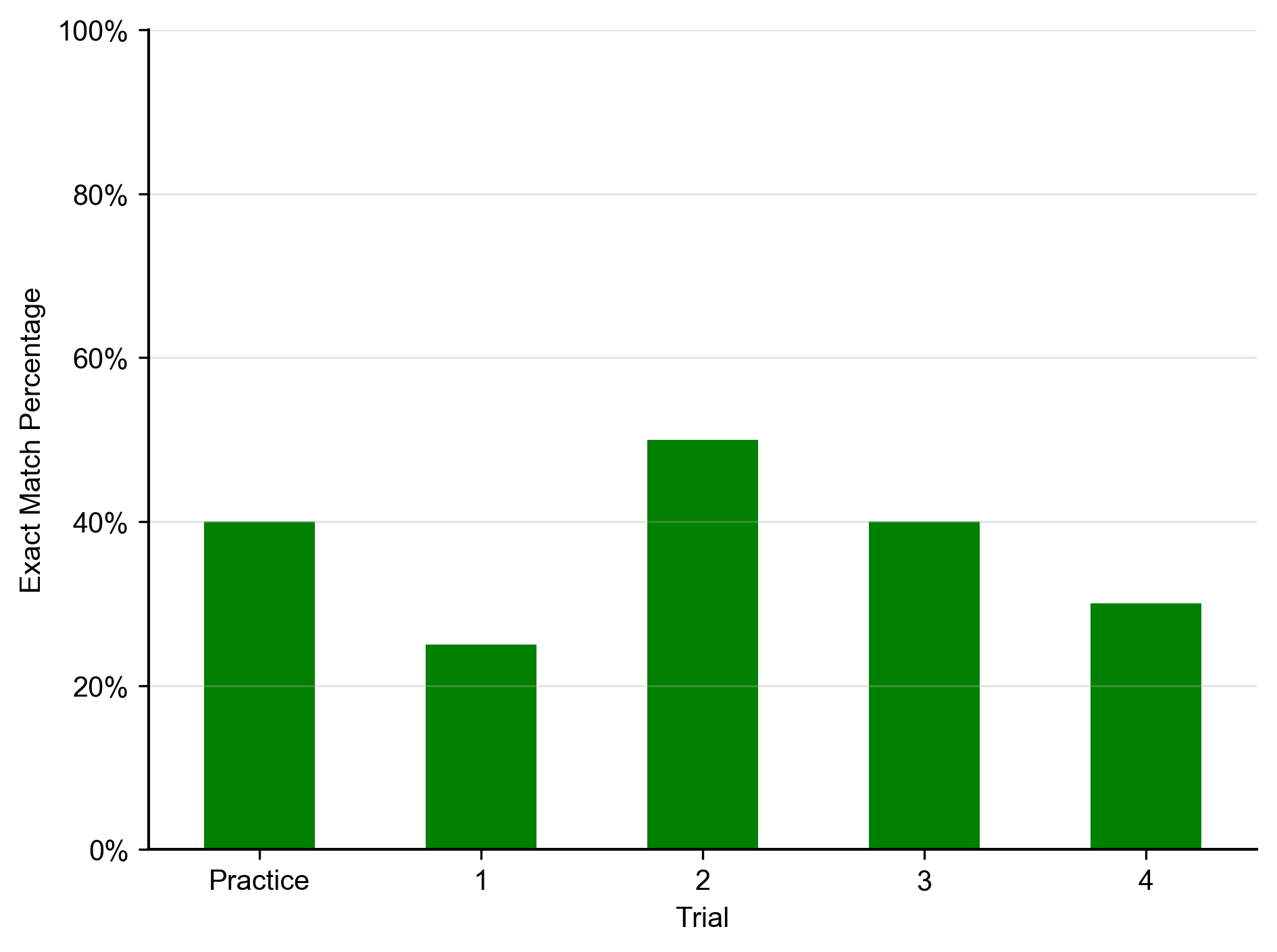}
\end{minipage}
\caption{Exact matches of the final puzzle configuration compared to target solution across conditions and trial. Left: the shared view condition, Right: the non-shared view condition.}
\label{fig:accuracies}
\end{figure}

Fig.~\ref{fig:accuracies} shows how often the pairs (human participants and AI) were able to create an exact match with the target solution. There are more exact matches when the view is shared than when it is not. The differences between the conditions are statistically significant ($\chi^2(1) = 11.57, p $<$ 0.001$). Within each condition, accuracy does not change significantly across trials, failing to demonstrate a learning effect. This confirms our expectation that in the non-shared view, the pairs will make more errors, but the lack of improvement over trials does not support our expectation of a learning effect.

\paragraph{Comparisons to expected observations on communication effort (Task-level)}
\begin{figure}[t]
    \centering
    \includegraphics[width=0.4\linewidth]{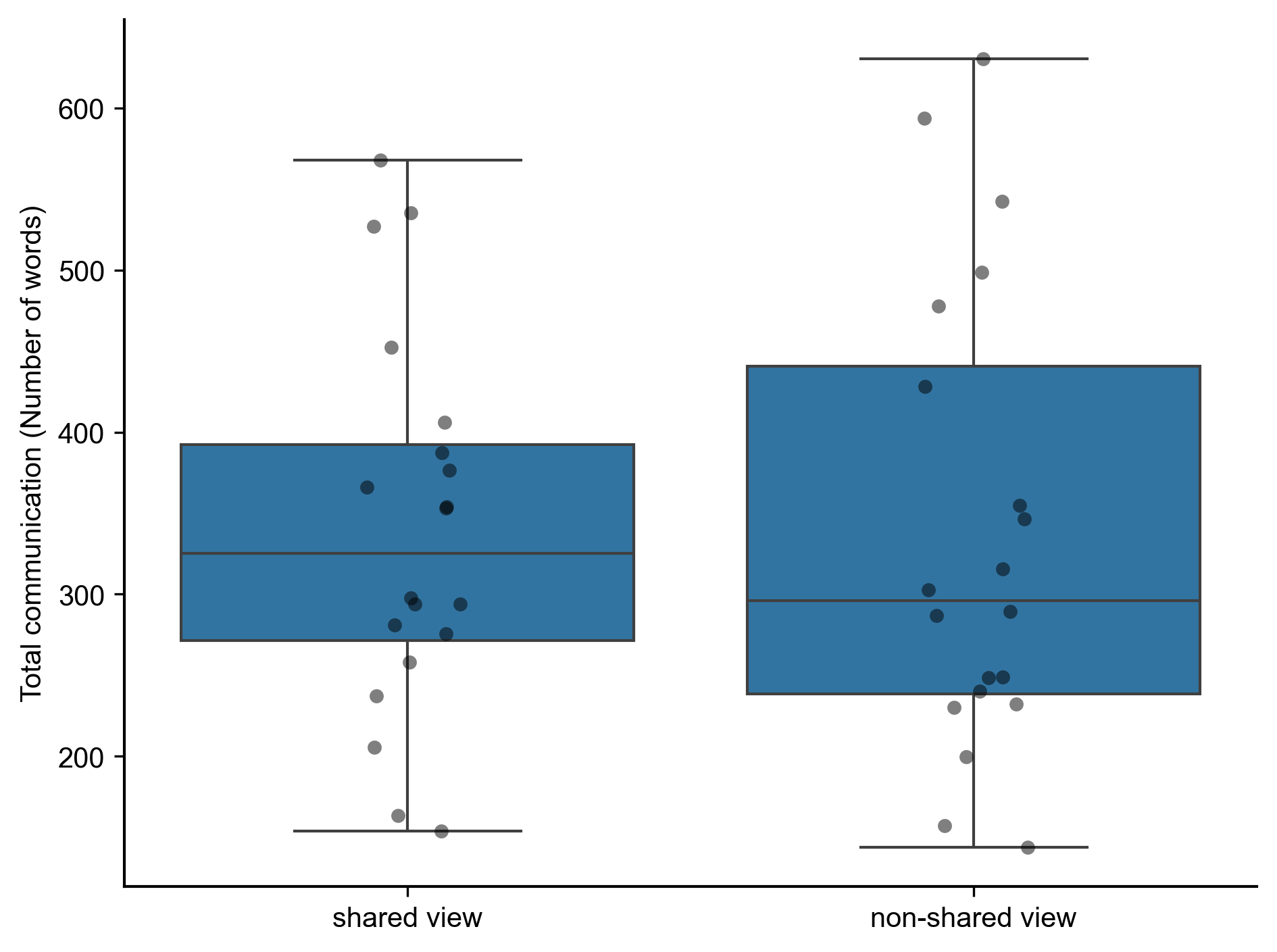}
    \caption{Total number of words used in the messages averaged across roles and across all puzzle trials in each condition (excluding the practice trial). Left: the shared view condition. Right: the non-shared view condition.}
    \label{fig:total_communication}
\end{figure}

The overall communication effort is shown in Fig.~\ref{fig:total_communication} as the distribution and average of the mean number of words used in all messages to solve the task over all trials (excluding the practice trial). On average, the amount of communication is not statistically significantly different between the shared view and non-shared view condition (U = 186.0, p = 0.65).

In contrast to our expectation, we observe no significant effect of the presence or absence of a shared view on the amount of communication. The message logs give some insight into why this might be. When a participant plays the role of Worker, actions are substituted for language, as previously found by \citet{10.1145/587078.587084}. In this case communication is minimal and predominantly takes the form of confirmations such as "done" and "ok", as visual feedback is used to resolve ambiguity. Prior work has shown that when there is no shared view, grounding happens via other media; in~\citep{Gergle01012013} this was expressed as an increase in communication. We do not observe such an increase and instead observe that brief confirmations are also given in the non-shared view condition (see Fig.~\ref{fig:h1_example2}).

This lack of an increase may also reflect some attempts at grounding made by participants in the role of Helper when there is no shared view. We observe a tendency for them to use very specific language, including use of coordinates and identifiers (e.g. "ID16"), as a means of minimising ambiguity and adopting terms they believed to be clearly understood by the model. The use of these emerged in part through interaction with the model; identifiers in particular were only ever introduced to the interaction by the model as only it had access to them through the code used to enable movement of puzzle pieces. However, once introduced they were often leveraged by human participants. This seems to be an attempt to counter the increased difficulty in building common ground when there is no shared view, whilst leveraging the reviewability of text-based communication (see Fig.~\ref{fig:h1_example1}). This aligns with \citet{brennan1996conceptual} that grounding costs influence specificity and that reduced situation awareness increases costs.

\begin{figure}[h!]  
\setlength{\tabcolsep}{6pt}    
\begin{tabular}{>{\bfseries}l p{10cm}}
AI & The board is currently empty and I have 24 unique patterned pieces available. Could you please tell me which piece to place first and at which position on the grid?  \\
\rowcolor{gray!20}  
Human & The piece has a clockwise spiral pattern. Starting from the centre of the tile. THe tile is placed in the top left location. Row 1, column 1.  \\ 
\end{tabular}  
\caption{Example of human Helper using concise, direct, and unambiguous language in the \textbf{non-shared view} condition.}
\label{fig:h1_example1}
\end{figure} 

\begin{figure}[h!]  
\setlength{\tabcolsep}{6pt}    
\begin{tabular}{>{\bfseries}l p{10cm}}
AI & Place the light pink piece with the black spiral pattern directly below the striped piece, in the second row, first column.  \\
\rowcolor{gray!20}  
Human & ok done  \\ 
\end{tabular}  
\caption{Example of human Worker's brief communication after taking an action in the \textbf{non-shared view} condition.}
\label{fig:h1_example2}
\end{figure}

\begin{figure}[t]  
\centering  
\begin{minipage}{0.48\textwidth}  
    \centering  
    \includegraphics[width=\textwidth]{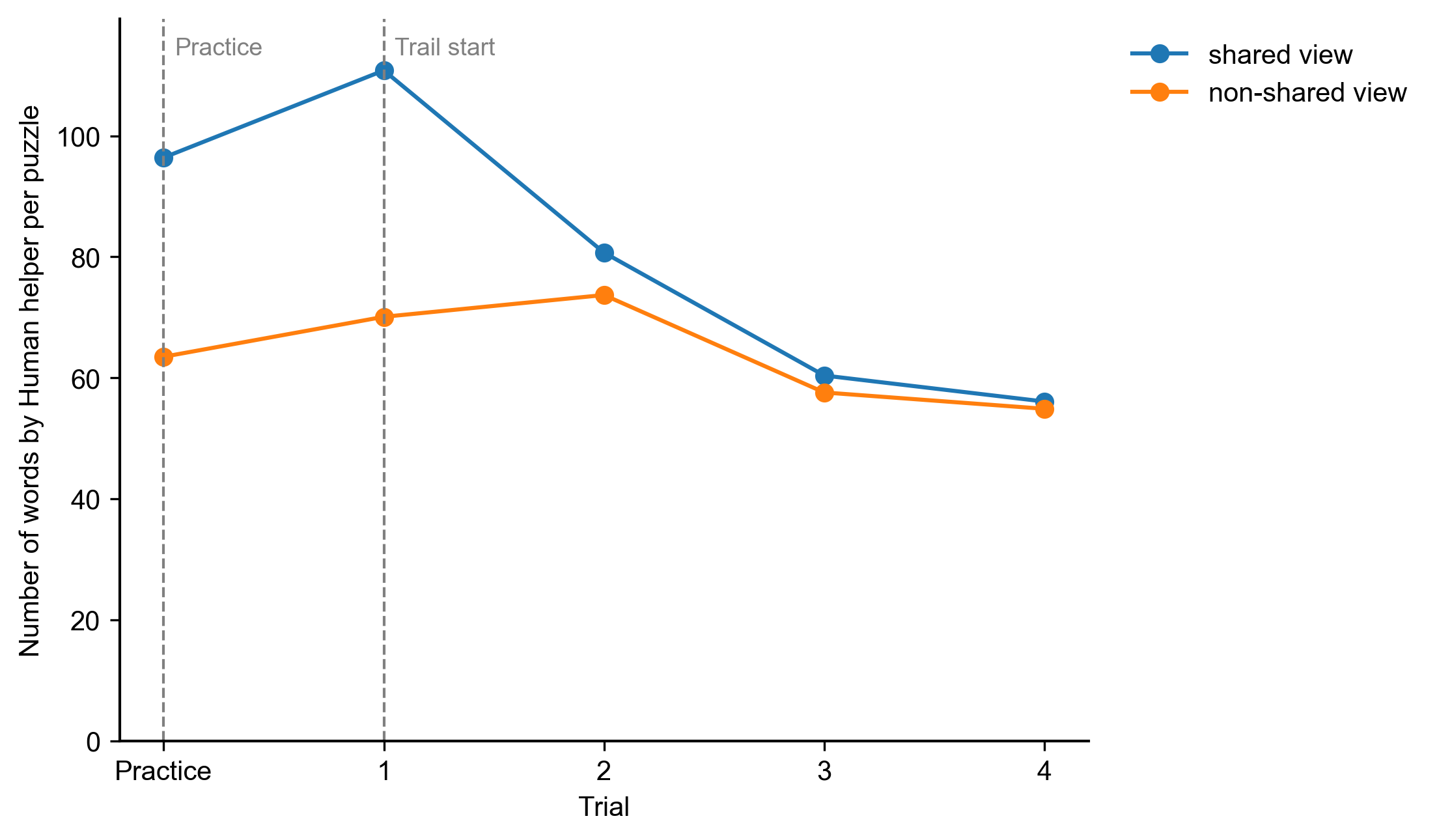}  
    \caption*{(a) Human Helper communication efforts as number of words used to solve each puzzle.}
\end{minipage}  
\hfill  
\begin{minipage}{0.48\textwidth}  
    \centering  
    \includegraphics[width=\textwidth]{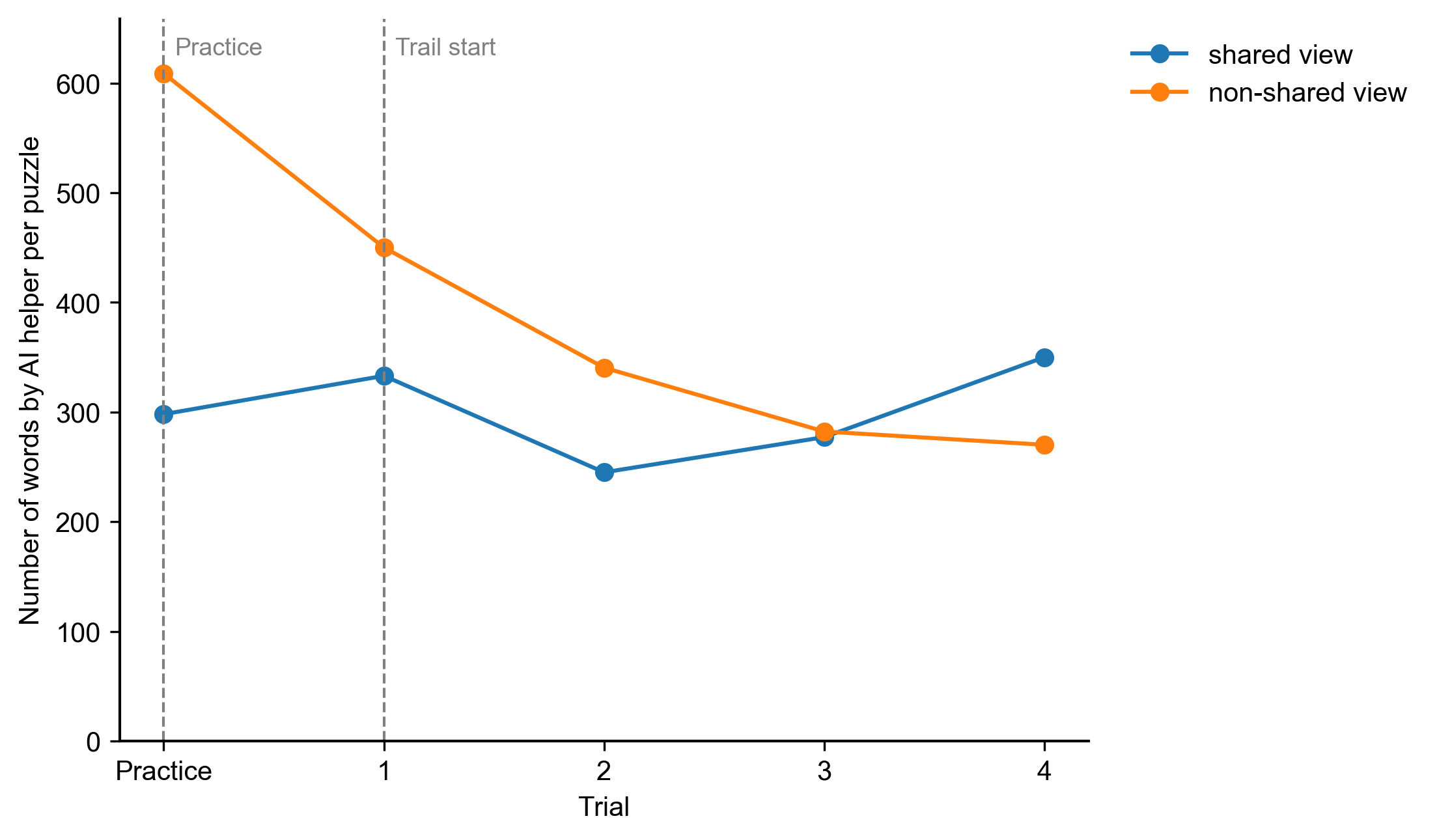}  
    \caption*{(b) AI Helper communication efforts as number of words used to solve each puzzle.} 
\end{minipage}  

\caption{Number of words in messages exchanged between participants and AI whilst solving each puzzle trial when the view is shared and not shared. (a) Number of words across all messages per puzzle trial, when the person is in the role of Helper. (b) Number of words across all messages per puzzle trial, when the person is in the role of Worker. The first trial served as a practice and is included for completeness.}  
\label{fig:human_ai_communication_2x2}  
\end{figure}  

Communication effort over the trials is illustrated in Fig.~\ref{fig:human_ai_communication_2x2}. Here, we focus on messages sent by Helpers only (for both human participants and AI). Communication efforts of Workers are low compared to Helpers, and are dependent on the Helper as they give the instructions. Consequently, previous studies~\citep{CLARK19861, Kraut2002Proximity, Gergle01012013} did not include the Worker in their analysis and concentrated on the Helper and total communications. Human Helper communication decreases significantly over the trials (F(1, 28) = 11.67, p $<$ 0.01) in the shared view condition, but not in the non-shared view condition. AI Helper communication does not decrease in the shared view condition, but does decrease significantly in the non-shared condition (F(1, 28) = 3.91, p $<$ 0.05). 

In accordance with our expectations, human Helpers become more efficient over time, at least when the view is shared. Further, reiterating our initial message log analysis above, the results in the non‑shared‑view condition might indicate that human Helpers adopt precise and unambiguous language early on, resulting in relatively stable communication effort across trials. The AI Helper shows a different pattern. In the shared view condition, AI communication remains largely unchanged over time, suggesting limited efficiency gains. In the non‑shared view condition, AI Helper communication decreases, but this reduction appears to be caused by a lack of engagement in grounding behaviours, such as asking for clarification.

This contrast highlights the importance of measures beyond surface‑level communication metrics and task completion. A benchmark is needed that enables evaluation of how people and AI coordinate, build up references, clarify, and resolve ambiguity. Understanding these interactions is crucial for understanding how and to what extent humans and AI build common ground. This is illustrated further in the following results.

\paragraph{Comparisons to expected observations referencing conventions (Object-level).}
If we examine noun phrases used to refer to puzzle pieces, we observe that the vast majority of these are used solely the by AI model. Noun phrases used solely by human participants, or that are jointly used by participants and AI, are fewer. Notably, the number of jointly used noun phrases decreases over trials in both conditions, but this trend is only statistical significant (F(1, 28) = 12.28, p $<$ 0.001) in the shared view condition. Further, the lengths of the noun phrases decrease over trials, but this effect is only statistical significant (F(1, 28) = 4.04, p $<$ 0.05) in the shared view condition. We did not observe a statistically significant difference in the length of noun phrases between conditions. 

While vocabulary that is jointly used is small compared to that used by the AI model, and decreasing over trials, it is interesting to note that the majority of noun phrases used by human participants are shared with the AI. The decrease in noun phrases used solely by participants, as well as the decrease in shared vocabulary, could indicate the use of a small set of relevant references over trials. In inspecting noun phrases in the messages logs, we see that the majority of differences in vocabulary between human and AI originates from the level of detail of descriptive references. AI uses lots of detail (see Fig.~\ref{fig:example2}) and repeats this over turns; e.g., repeating "the pink piece with a checkerboard pattern", rather than moving to the more efficient "the pink checkerboard", or "the checkerboard" once ambiguity has been removed. In contrast, people use less detailed descriptions (see Fig.~\ref{fig:example1}). 

\begin{figure}[h!]  
\setlength{\tabcolsep}{6pt}    
\begin{tabular}{>{\bfseries}l p{10cm}}  
\rowcolor{gray!20}  
Human & spiral in top left  \\ 
AI & Great, the spiral is now at the top left (0,0). What piece and position should I place next? Please describe the pattern or color and the location.  \\
\rowcolor{gray!20}  
Human & place the checkerboard to the right of it  \\ 
\end{tabular}  
\caption{Example of referential language used by human helper in the \textbf{shared view} condition.}
\label{fig:example1}
\end{figure} 

\begin{figure}[h!]  
\setlength{\tabcolsep}{6pt}    
\begin{tabular}{>{\bfseries}l p{10cm}}  
\rowcolor{gray!20}  
AI & Place the pink piece with a spiral pattern at the top left corner of the grid.  \\
Human & ok next  \\ 
\rowcolor{gray!20}  
AI & Great start! Now, place the white piece with a black crisscross diamond pattern directly to the right of the pink spiral piece, at the top row, second column..  \\
\end{tabular}  
\caption{Example of referential language used by AI helper in the \textbf{shared view} condition.}
\label{fig:example2}
\end{figure} 

While this does not confirm our expectation of the establishment of shared referential conventions, many of the descriptive references may still be clear in terms of their referential purpose, even when they differ lexically. Inspecting the messages logs further, we do observe that the references to the pieces are mostly definite, indicating an assumption that they are understood by their collaborative partner. These patterns show the importance of measuring assumptions about mutual understanding of the references. A benchmark is needed that enables the assessment of the intentions of the references by looking at their types, in addition to lexical convergence.

\begin{figure}[t]  
\centering  
\begin{minipage}{0.48\textwidth}  
    \centering  
    \includegraphics[width=\textwidth]{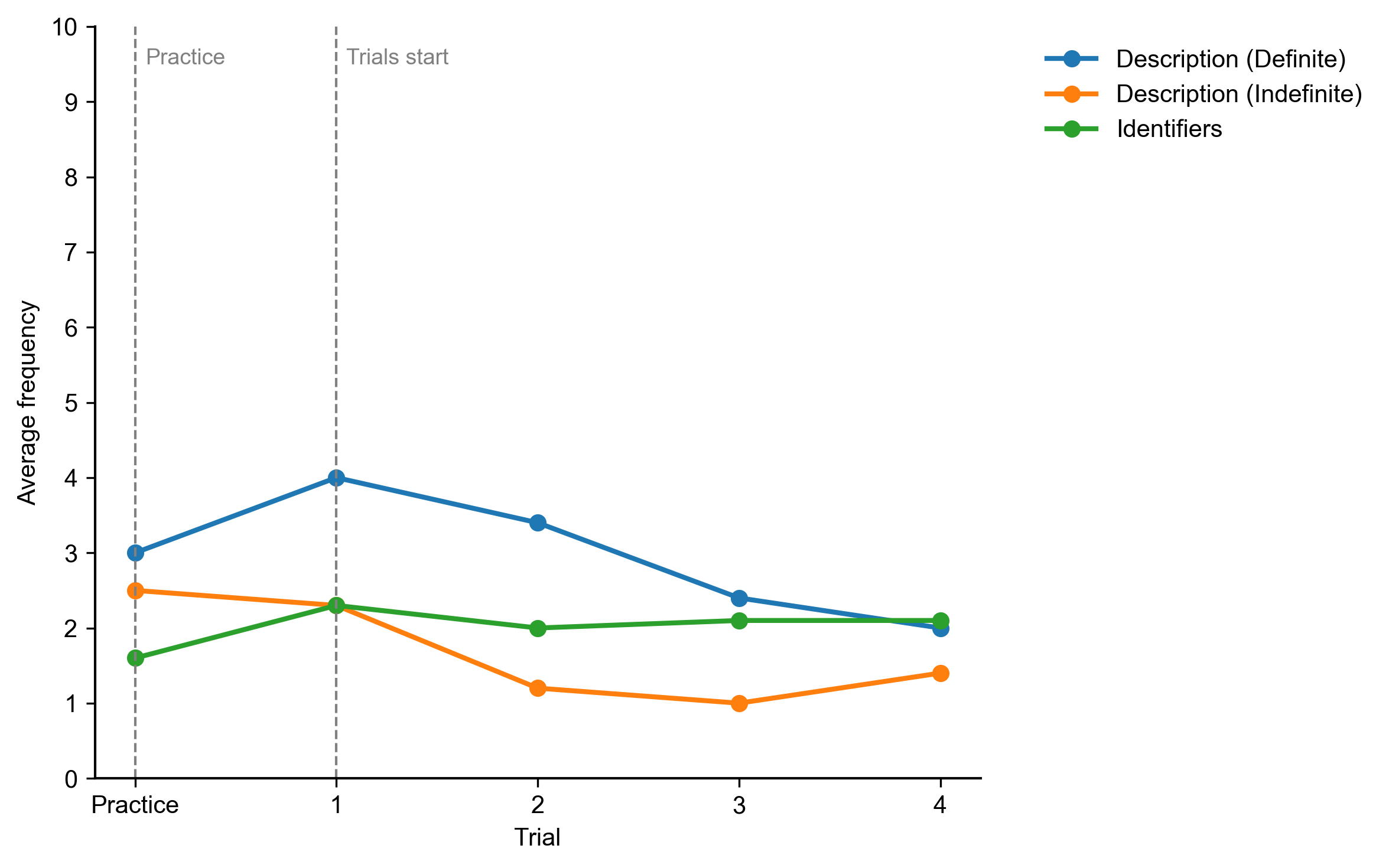}  
    \caption*{(a) Human helper, shared view condition.}
\end{minipage}  
\hfill  
\begin{minipage}{0.48\textwidth}  
    \centering  
    \includegraphics[width=\textwidth]{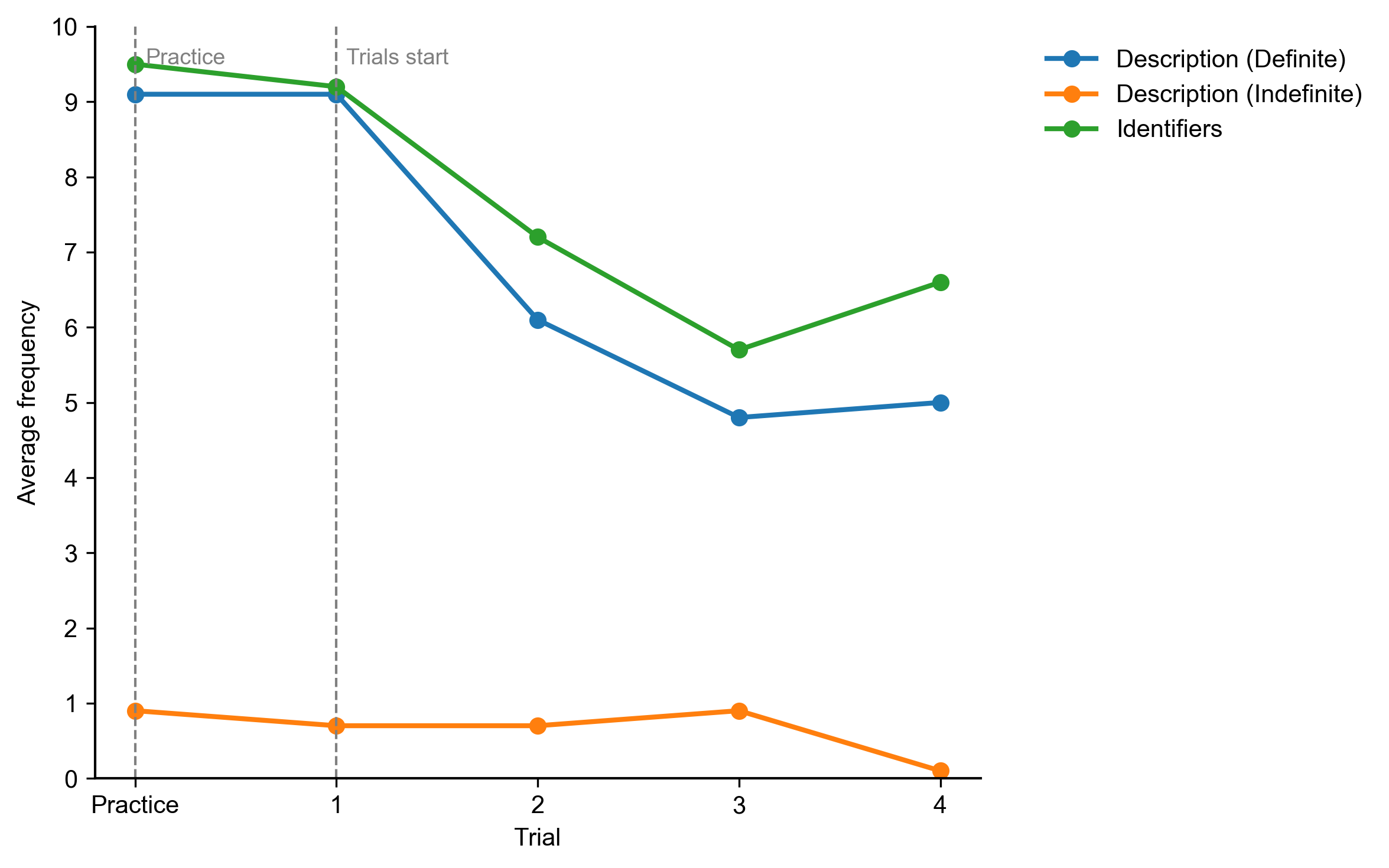}  
    \caption*{(b) AI worker, shared view condition.}
\end{minipage}  
  
  
\begin{minipage}{0.48\textwidth}  
    \centering  
    \includegraphics[width=\textwidth]{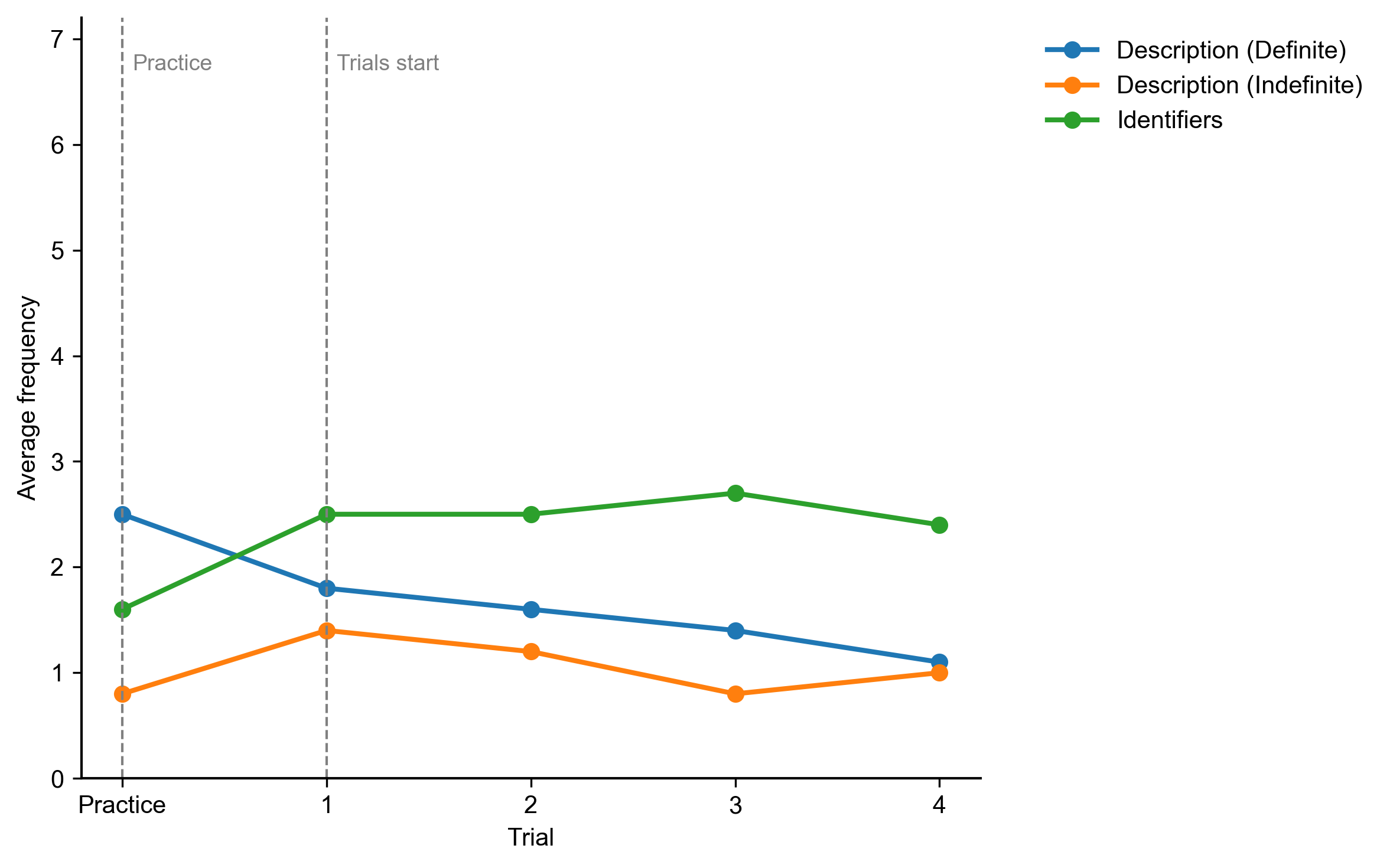}  
    \caption*{(c) Human helper, non-shared view condition.}
\end{minipage}  
\hfill  
\begin{minipage}{0.48\textwidth}  
    \centering  
    \includegraphics[width=\textwidth]{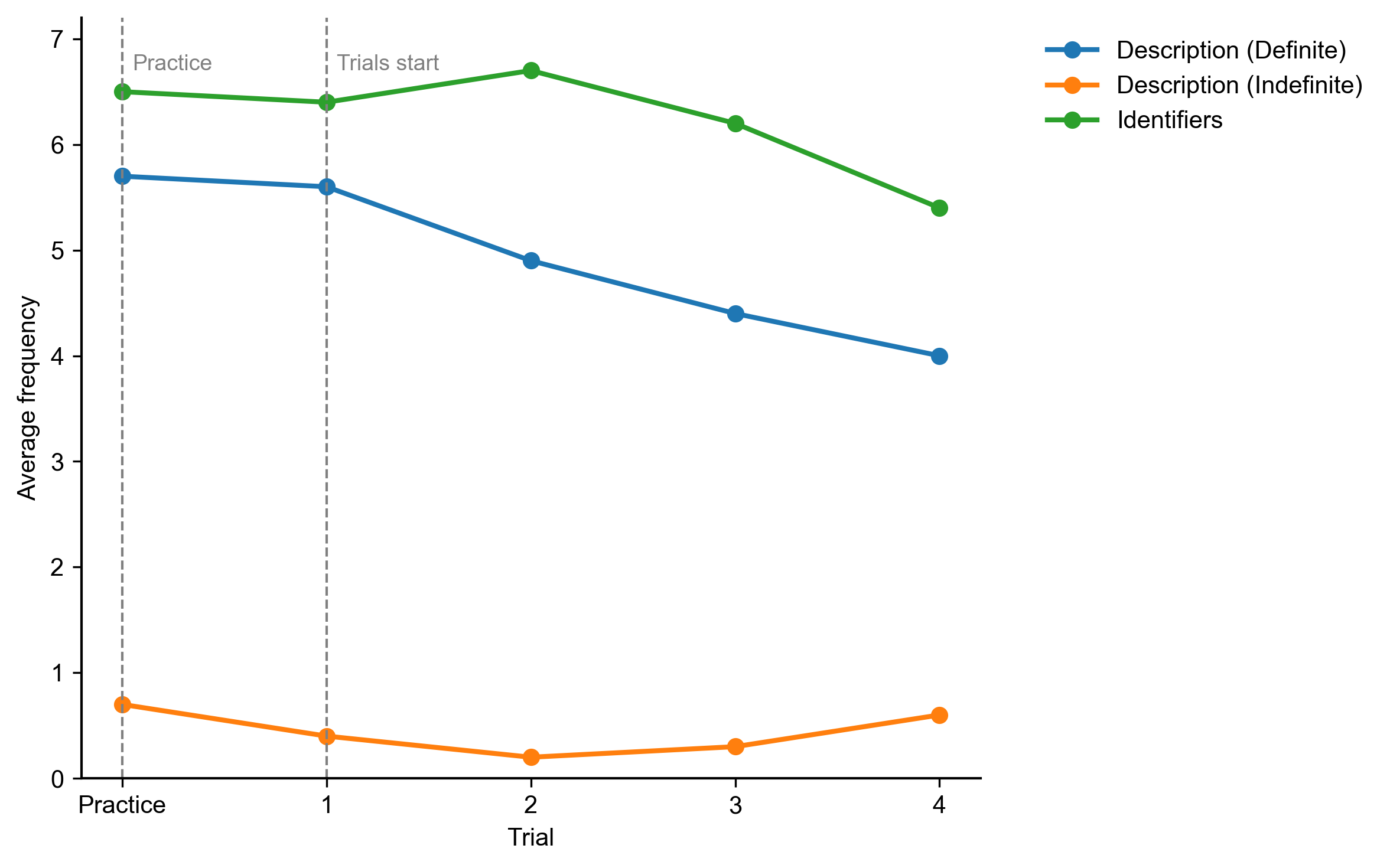}  
    \caption*{(d) AI workers, non-shared view condition.}  
\end{minipage}  
  
\caption{Usage of reference types (descriptions vs. identifiers) across puzzle trials for human Helpers and AI Workers in the shared and non-shared view conditions.}  
\label{fig:reference_types_human_helper_shared_vs_nonshared_2x2}
\end{figure}

\begin{figure}[t]  
\centering  
\begin{minipage}{0.48\textwidth}  
    \centering  
    \includegraphics[width=\textwidth]{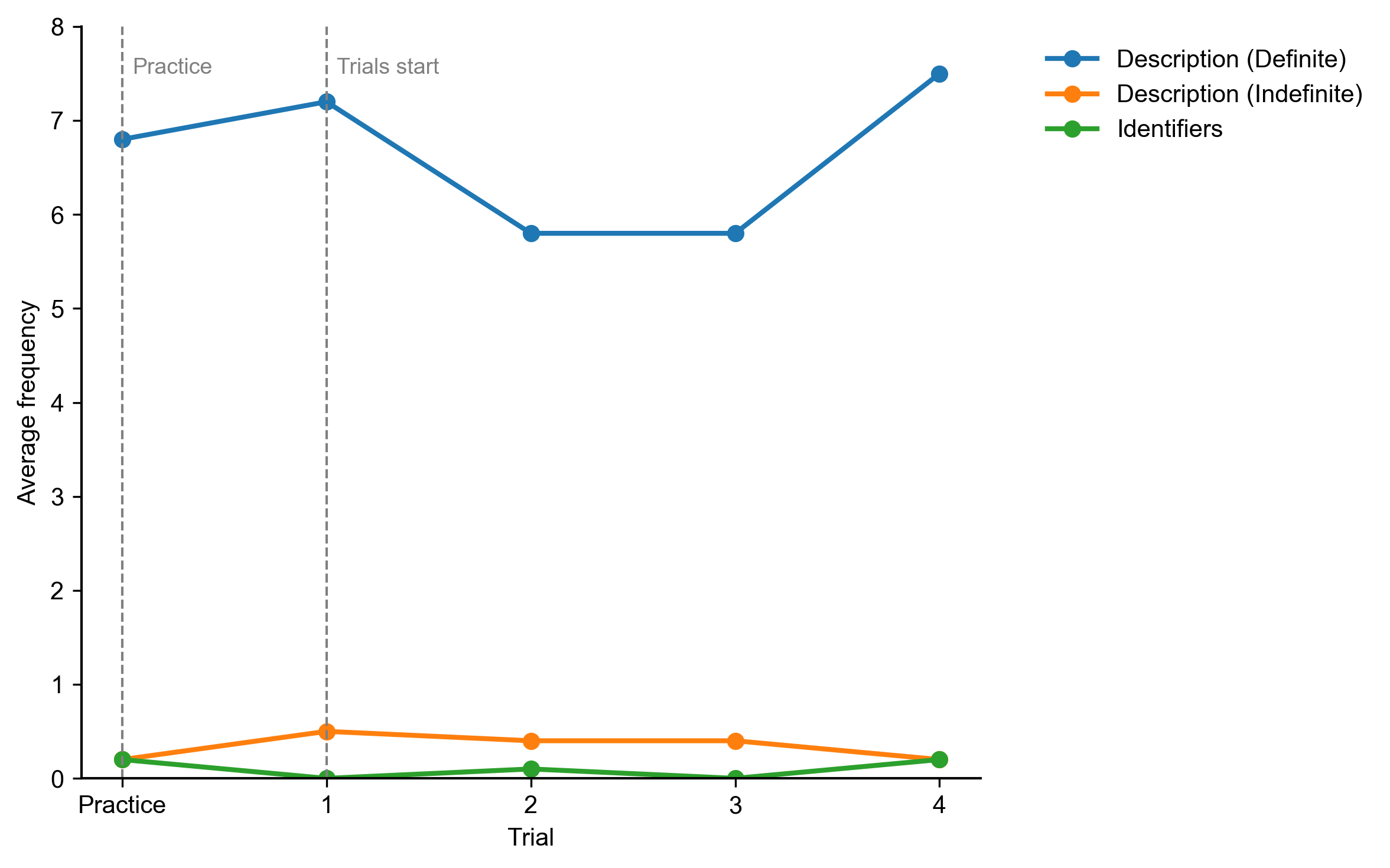}  
    \caption*{(a) AI Helper, shared view condition.\\}
\end{minipage}  
\hfill  
\begin{minipage}{0.48\textwidth}  
    \centering  
    \includegraphics[width=\textwidth]{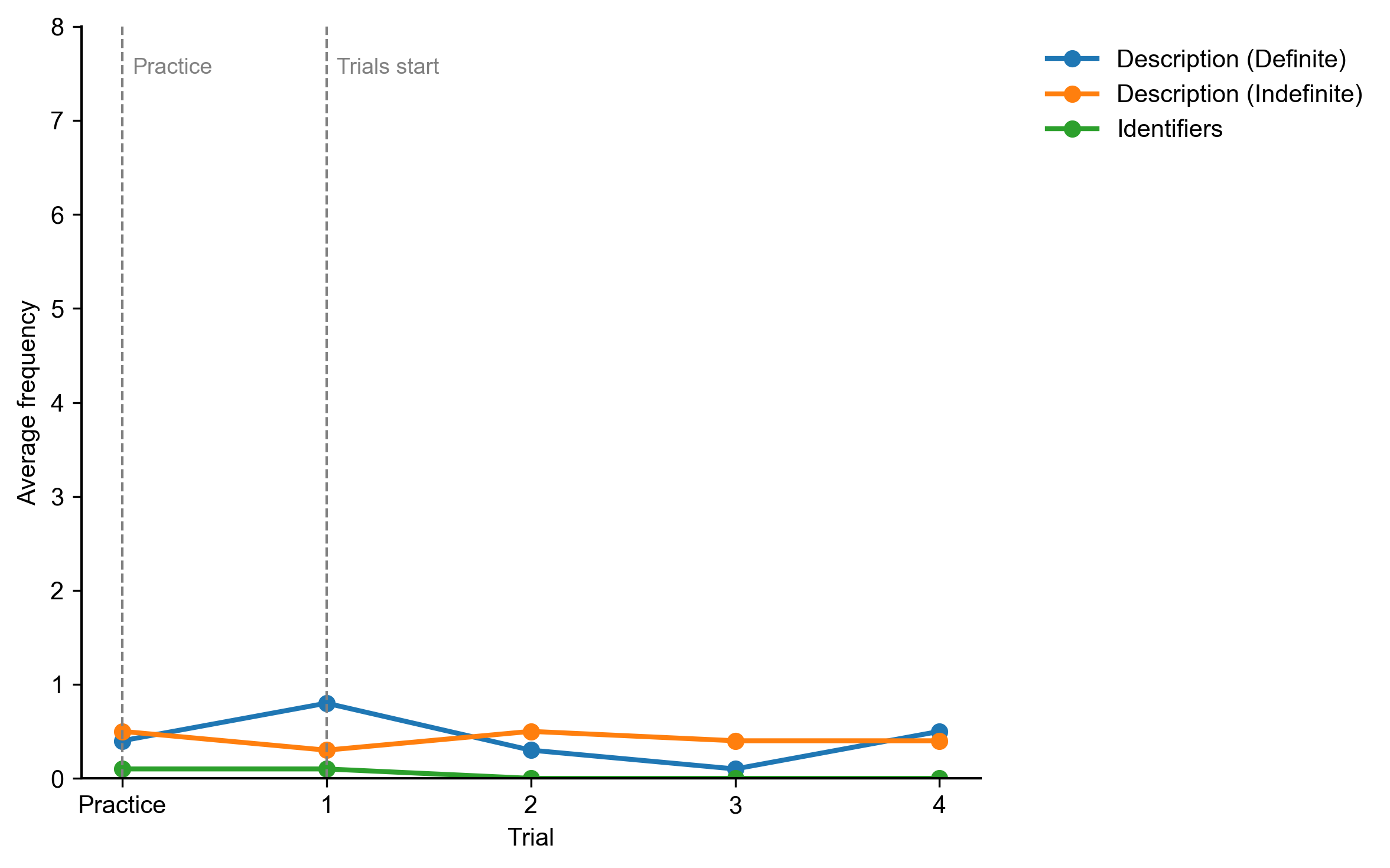}  
    \caption*{(b) Human Worker, shared view condition.}
\end{minipage}  
  
  
\begin{minipage}{0.48\textwidth}  
    \centering  
    \includegraphics[width=\textwidth]{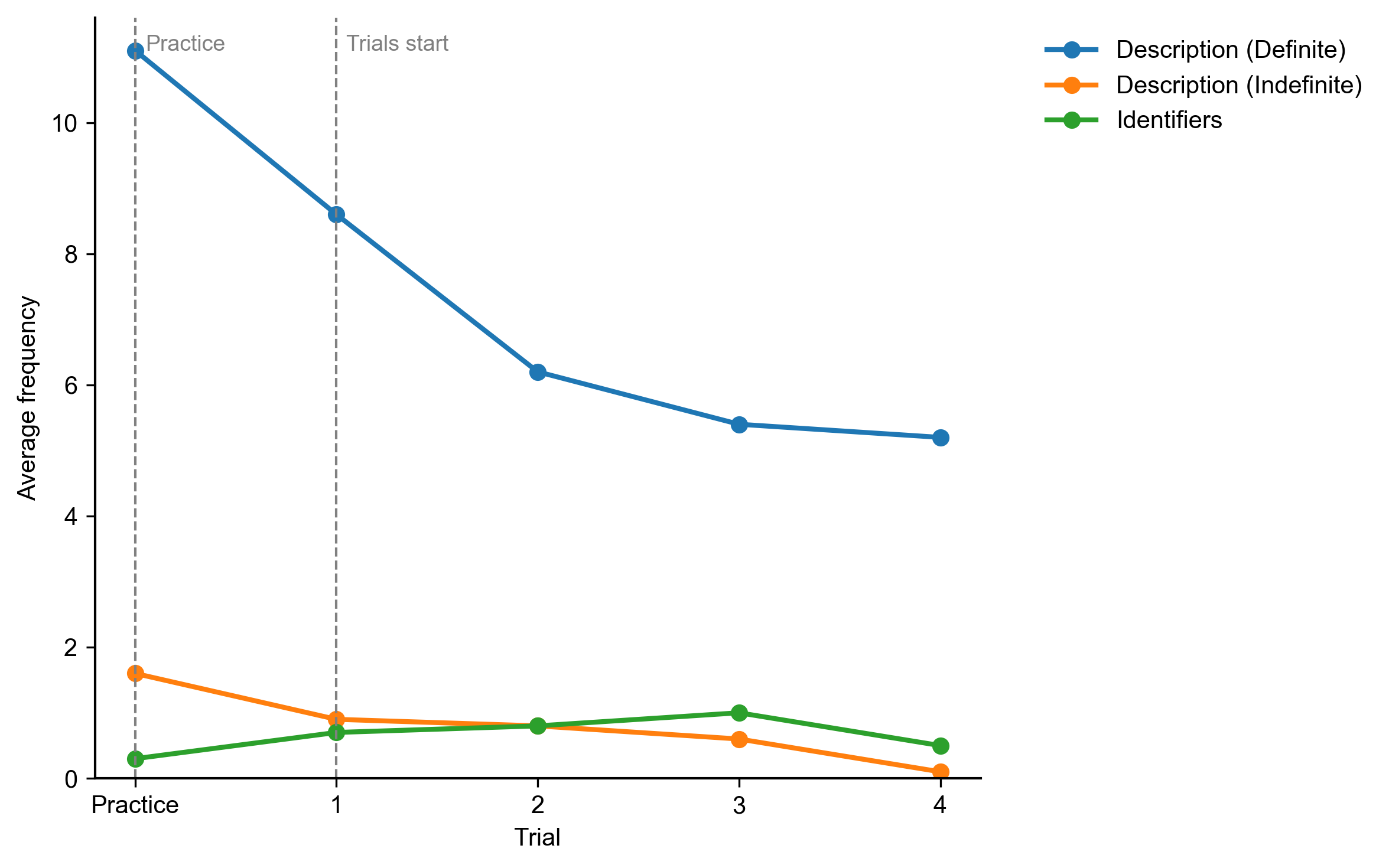}  
    \caption*{(c) AI Helper, non-shared view condition.\\}
\end{minipage}  
\hfill  
\begin{minipage}{0.48\textwidth}  
    \centering  
    \includegraphics[width=\textwidth]{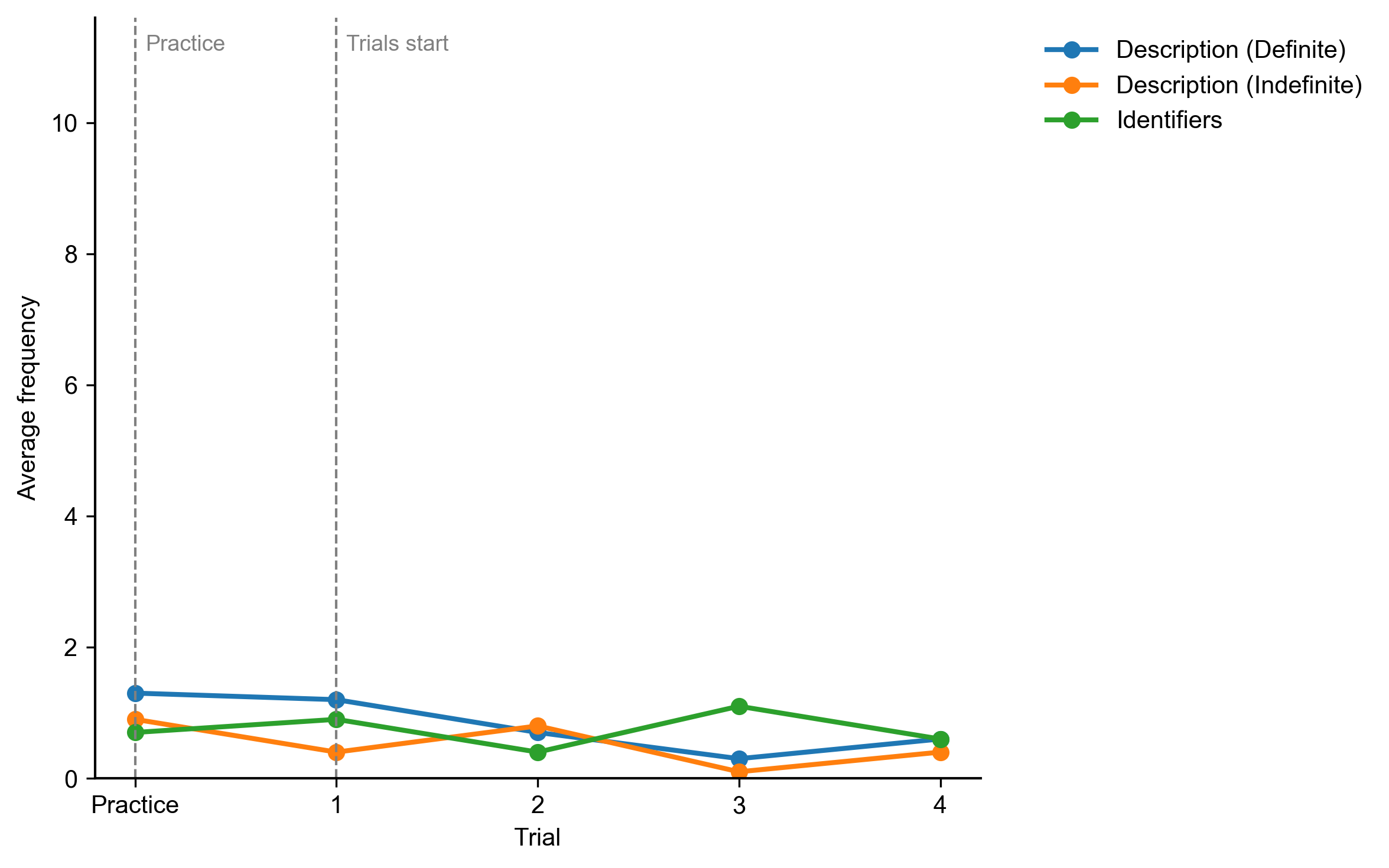}  
    \caption*{(d) Human Worker, non-shared view condition.}  
\end{minipage}  
  
\caption{Usage of reference types (descriptions vs. identifiers) across puzzle trials for AI Helpers and human Workers in the shared and non-shared view conditions.}  
\label{fig:reference_types_AI_helper_shared_vs_nonshared_2x2}
\end{figure} 

Fig.~\ref{fig:reference_types_human_helper_shared_vs_nonshared_2x2} illustrates the different types of references used by human Helpers and AI Workers across the trials and conditions. Use of descriptive references reduces over trials, with definite descriptions reducing significantly in the shared view condition (F(1, 28) = 11.64 p $<$ 0.001). Use of identifiers shows a trend to increase when there is no shared view, but the difference does not reach significance. The AI Worker uses both identifiers and descriptions at a high rate, but reduces significantly over trials for definite descriptions (F(1, 28) = 13.92, p $<$ 0.0001) and for identifiers (F(1, 28) = 7.87, p $<$ 0.01). In contrast, the human participant Worker~\ref{fig:reference_types_AI_helper_shared_vs_nonshared_2x2} rarely uses references across all trials. The AI Helper uses a high rate of definite descriptions; this reduces significantly over trials when there is no shared view (F(1, 28) = 5.06 p $<$ 0.05), but even at its lowest level it is higher than the human helper.

Given that we observed only a small joint vocabulary within pairs to refer to puzzle pieces, we compare referencing conventions for people and AI separately. First, in contrast to \citet{hh1981definite}, we observe that both people and AI use definite references from the early trials. This may in part reflect the revisability and reviewability of text; it may be easier to produce descriptive references through writing, and to understand them through reading, than in prior studies of spoken communication. Furthermore, when the view is shared, mistakes are easily spotted and can be repaired. In contrast to results reported by \citet{fussell1992coordination}, participants and AI might use definite references even if their partner may not understand them, while relying on the virtual co-presence~\citep{10.1145/587078.587084} and visual feedback for repairs. 

Despite the differences between how people and AI refer to puzzle pieces, we do observe a convergence towards a shared referential convention, in line with our expectations. As noted above, people adapted to using identifier-based references, which were introduced by the AI Worker (see Fig.~\ref{fig:example_ref_2}). Identifiers emerge because the AI worker generates actions using a domain-specific-language (DSL) that requires explicit identifiers and coordinates for function calls. While very specific to human-AI interaction in this condition, this can also be leveraged as a mechanism to examine whether referential conventions proposed by the AI are adopted by people.

\begin{figure}[h!]  
\setlength{\tabcolsep}{6pt}    
\begin{tabular}{>{\bfseries}l p{10cm}}
AI & I've placed the yellow checkerboard pattern piece (ID 10) at position (1, 0), directly to the right of the spiral. What should I do next?  \\
\rowcolor{gray!20}  
Human & put ID0 on 0,0  \\ 
\end{tabular}  
\caption{Example of human helper using the identifiers as references in the \textbf{non-shared view} condition.}
\label{fig:example_ref_2}
\end{figure} 

Identifiers are concise and offer a way for people to minimise typing and communicative effort. However, it is not clear whether their use supports the principle of least collaborative effort~\citep{CLARK19861}. The AI Worker uses them alongside descriptive references, so reading is not reduced. Further, while the reviewability of text means that identifiers can be cross-checked within a trial, remembering identifiers across trials adds to the participant's cognitive load. This can be mitigated by asking the AI to list identifiers and pieces, and in some cases, participants did just that. But in other cases, participants bore the cognitive effort of remembering identifiers. Either way, use of identifiers seem to reflect a convention pact~\citep{brennan1996conceptual} in which people adapted to the AI models way of communicating, possibly as a means of assuring its understanding of their messages to it. 

Despite this (asymmetric) convergence in referential practices and the high usage of definite references, it remains unclear whether mutual understanding, and common ground, is really being built. To assess this, a benchmark must reveal how references are introduced, evaluated through clarification, repaired if necessary, and ultimately accepted through interaction. Therefore, in the next section we show the analysis of conversational grounding behaviors.

\paragraph{Comparisons to expected observations on conversational grounding (Communication-level)}

\begin{figure}[t]  
\centering  
\begin{minipage}{0.48\textwidth}  
    \centering  
    \includegraphics[width=\textwidth]{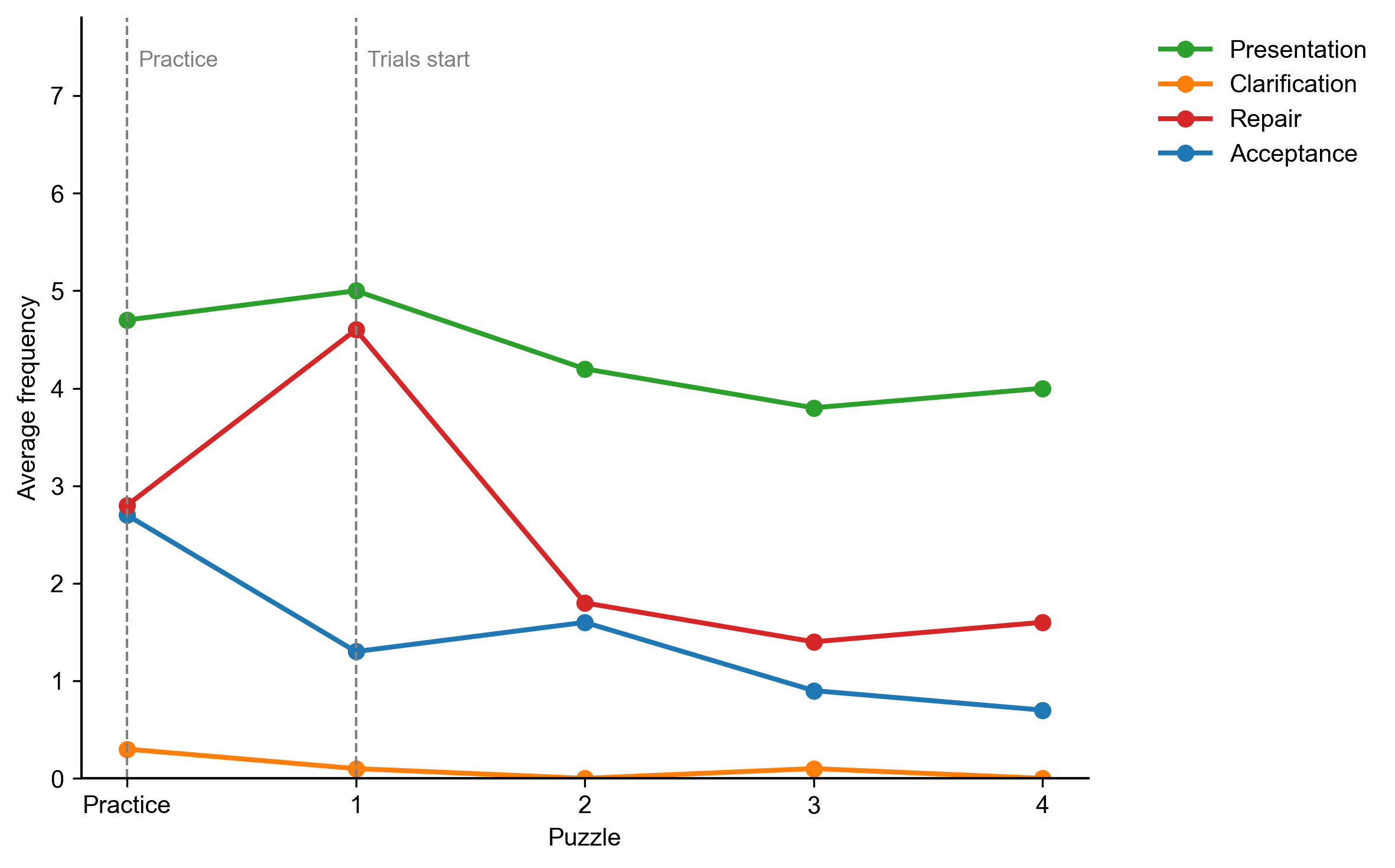}  
    \caption*{(a) Human Helper, shared view condition.}
\end{minipage}  
\hfill  
\begin{minipage}{0.48\textwidth}  
    \centering  
    \includegraphics[width=\textwidth]{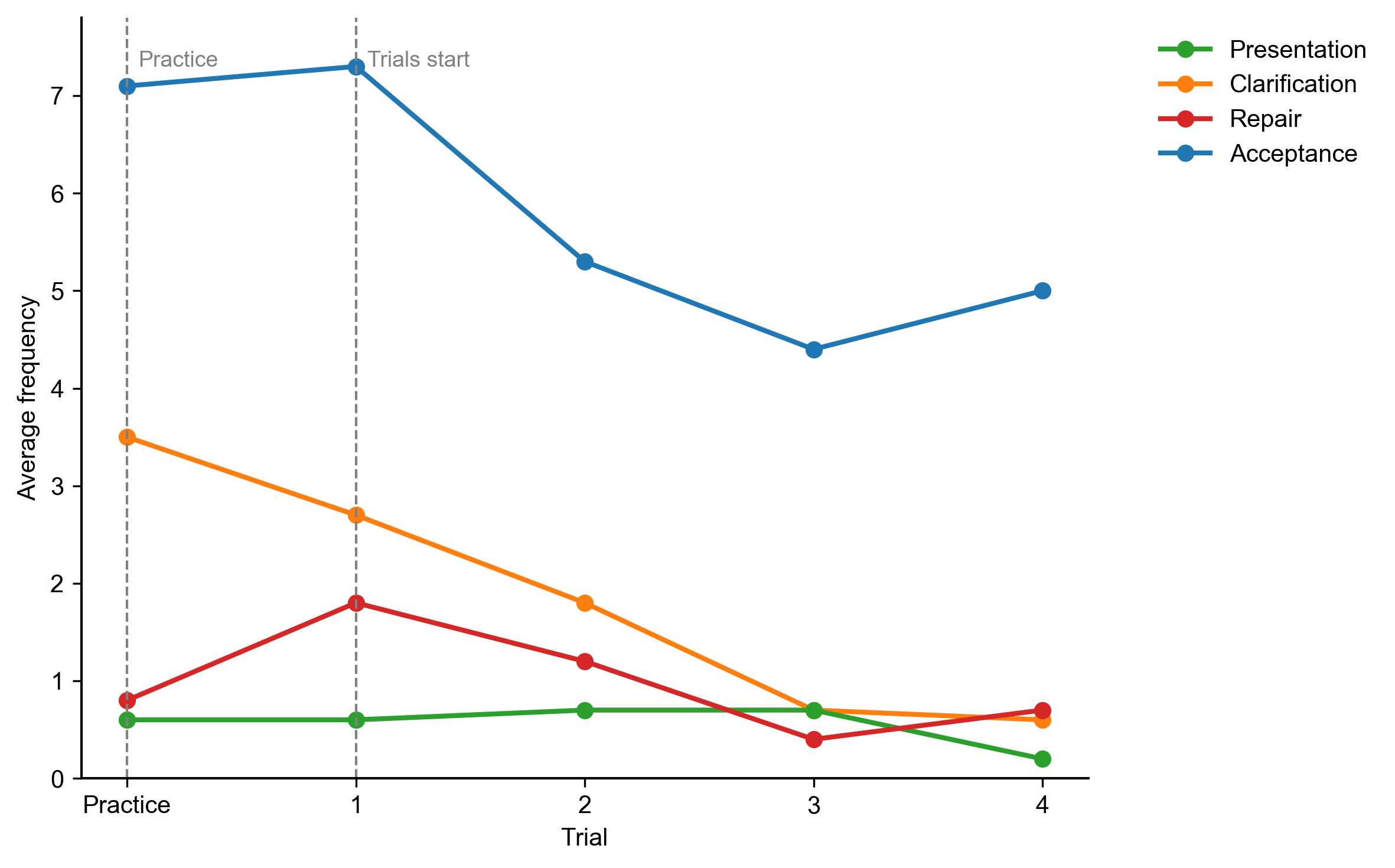}  
    \caption*{(b) AI Worker, shared view condition.\\}
\end{minipage}  
  
  
\begin{minipage}{0.48\textwidth}  
    \centering  
    \includegraphics[width=\textwidth]{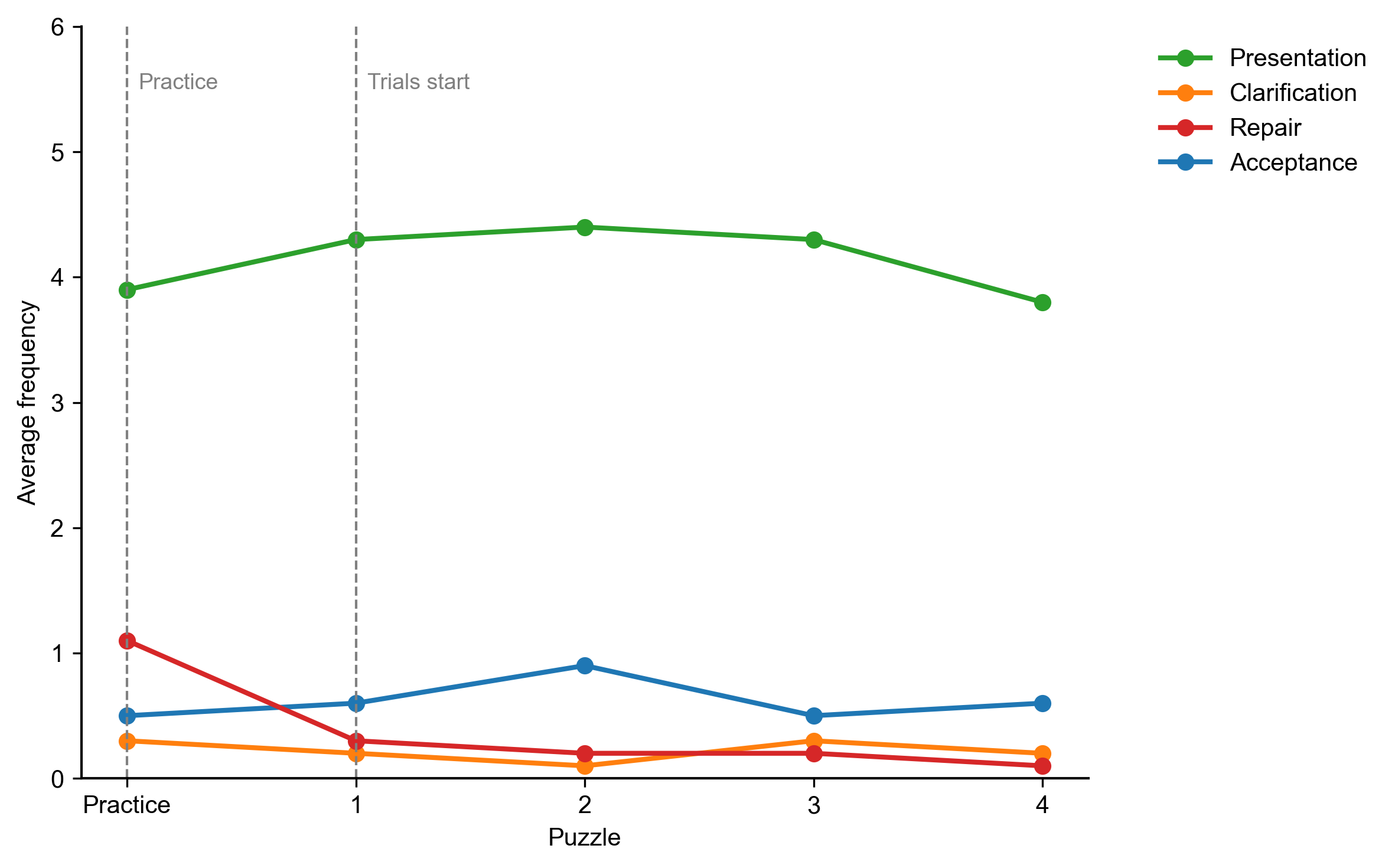}  
    \caption*{(c) Human Helper, non-shared view condition.\\}
\end{minipage}  
\hfill  
\begin{minipage}{0.48\textwidth}  
    \centering  
    \includegraphics[width=\textwidth]{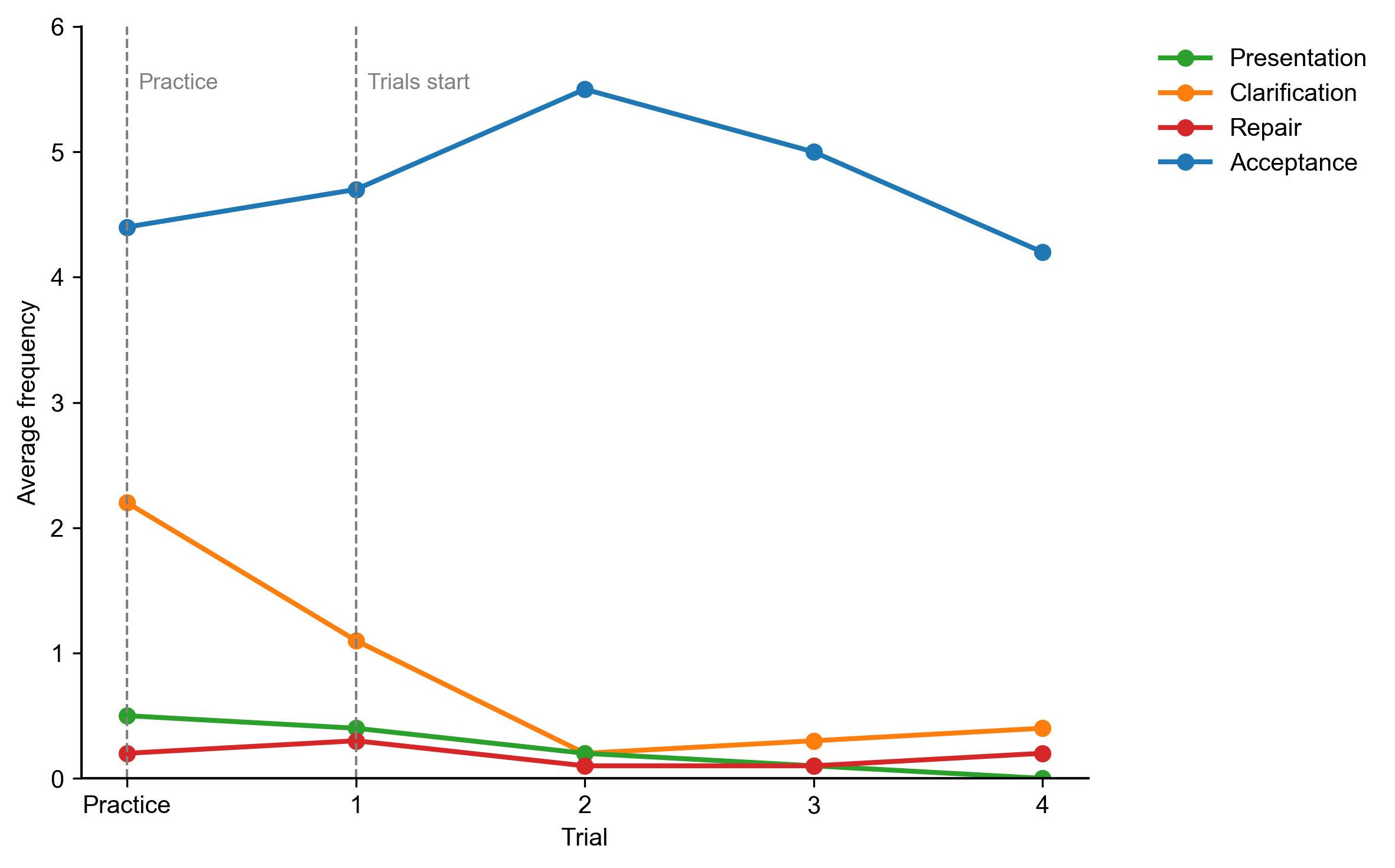}  
    \caption*{(d) AI Worker, non-shared view condition.}  
\end{minipage}  
  
\caption{Use of grounding acts (presentations, clarifications, repairs, and acceptances) in messages across trials, sent by people and by the AI model, in the shared and non-shared view conditions.}  
\label{fig:grounding_types_human_helper_shared_vs_nonshared_2x2}
\end{figure}

\begin{figure}[t]  
\centering  

\begin{minipage}{0.48\textwidth}  
    \centering  
    \includegraphics[width=\textwidth]{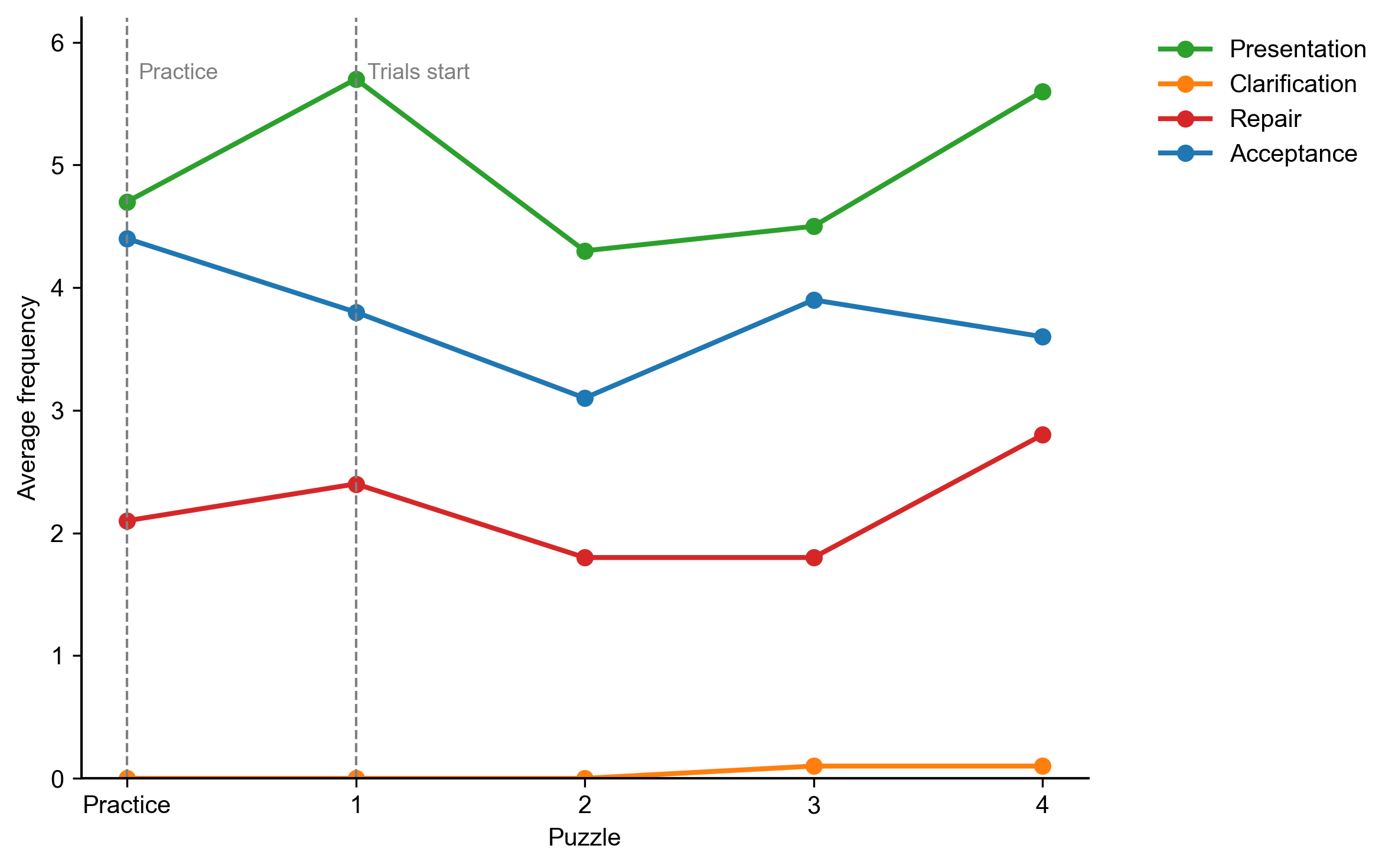}  
    \caption*{(a) AI helper, shared view condition.\\}
\end{minipage}  
\hfill  
\begin{minipage}{0.48\textwidth}  
    \centering  
    \includegraphics[width=\textwidth]{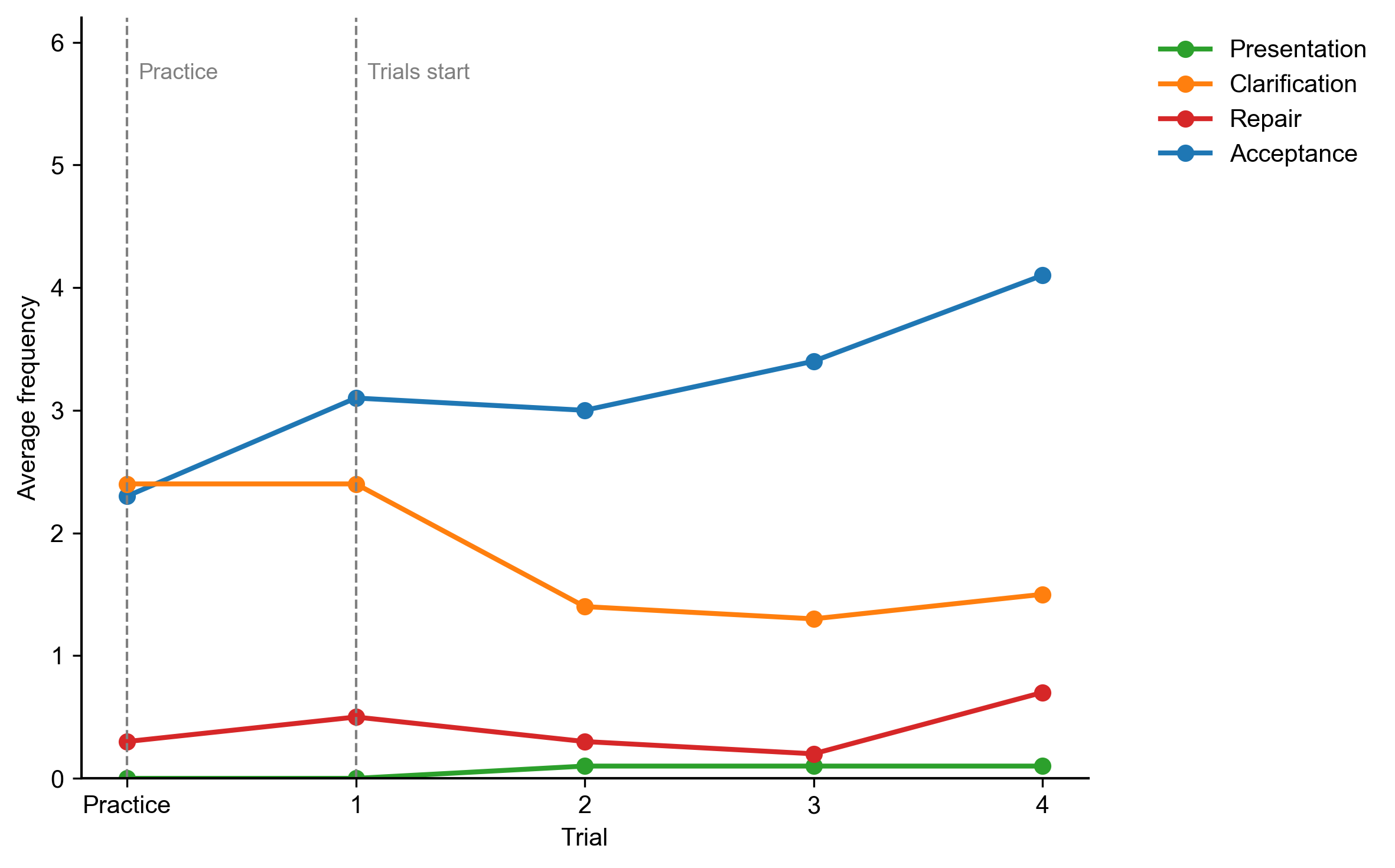}  
    \caption*{(b) Human worker, shared view condition.}
\end{minipage}  
  

\begin{minipage}{0.48\textwidth}  
    \centering  
    \includegraphics[width=\textwidth]{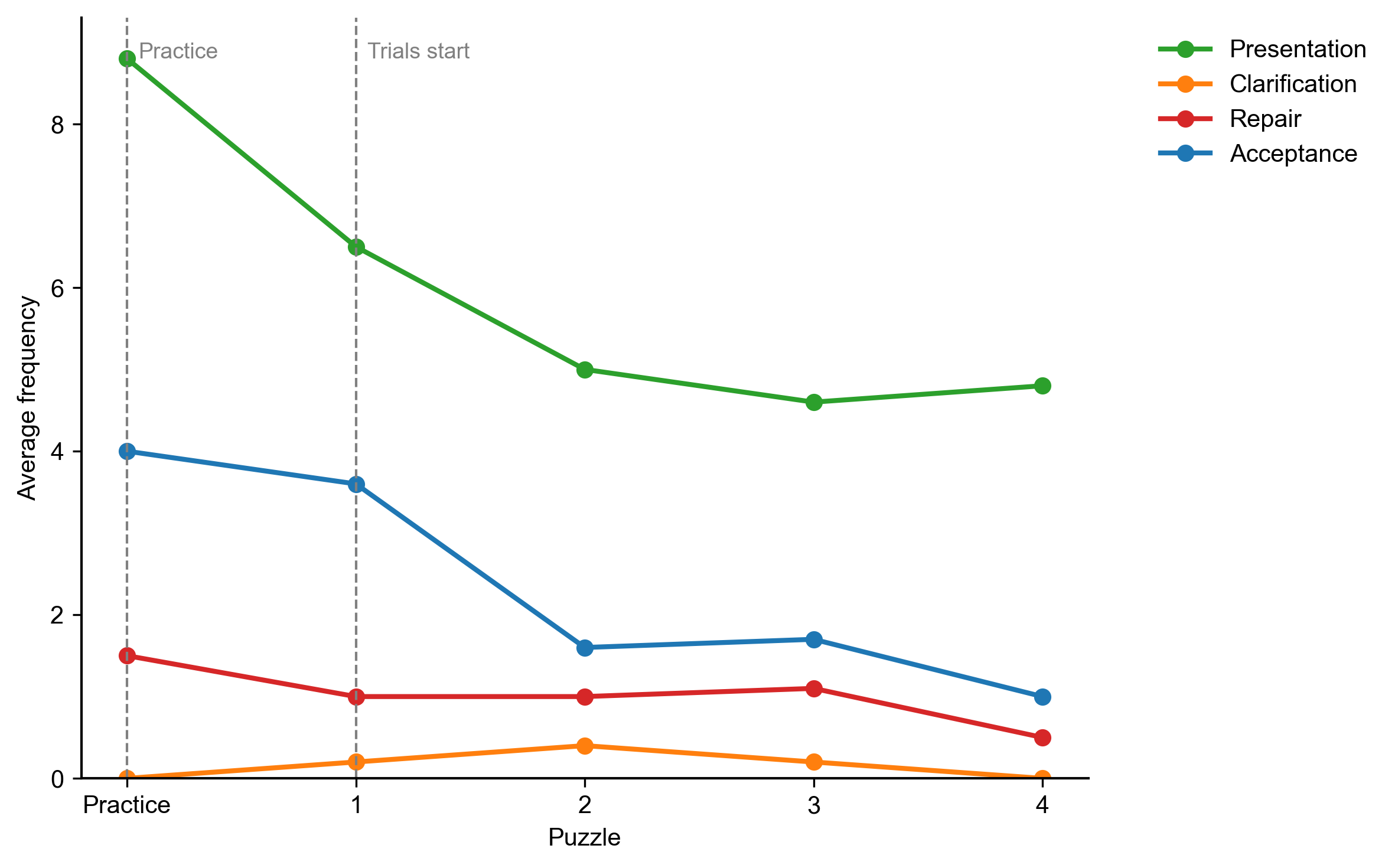}  
    \caption*{(c) AI helper, non-shared view condition.\\}
\end{minipage}  
\hfill  
\begin{minipage}{0.48\textwidth}  
    \centering  
    \includegraphics[width=\textwidth]{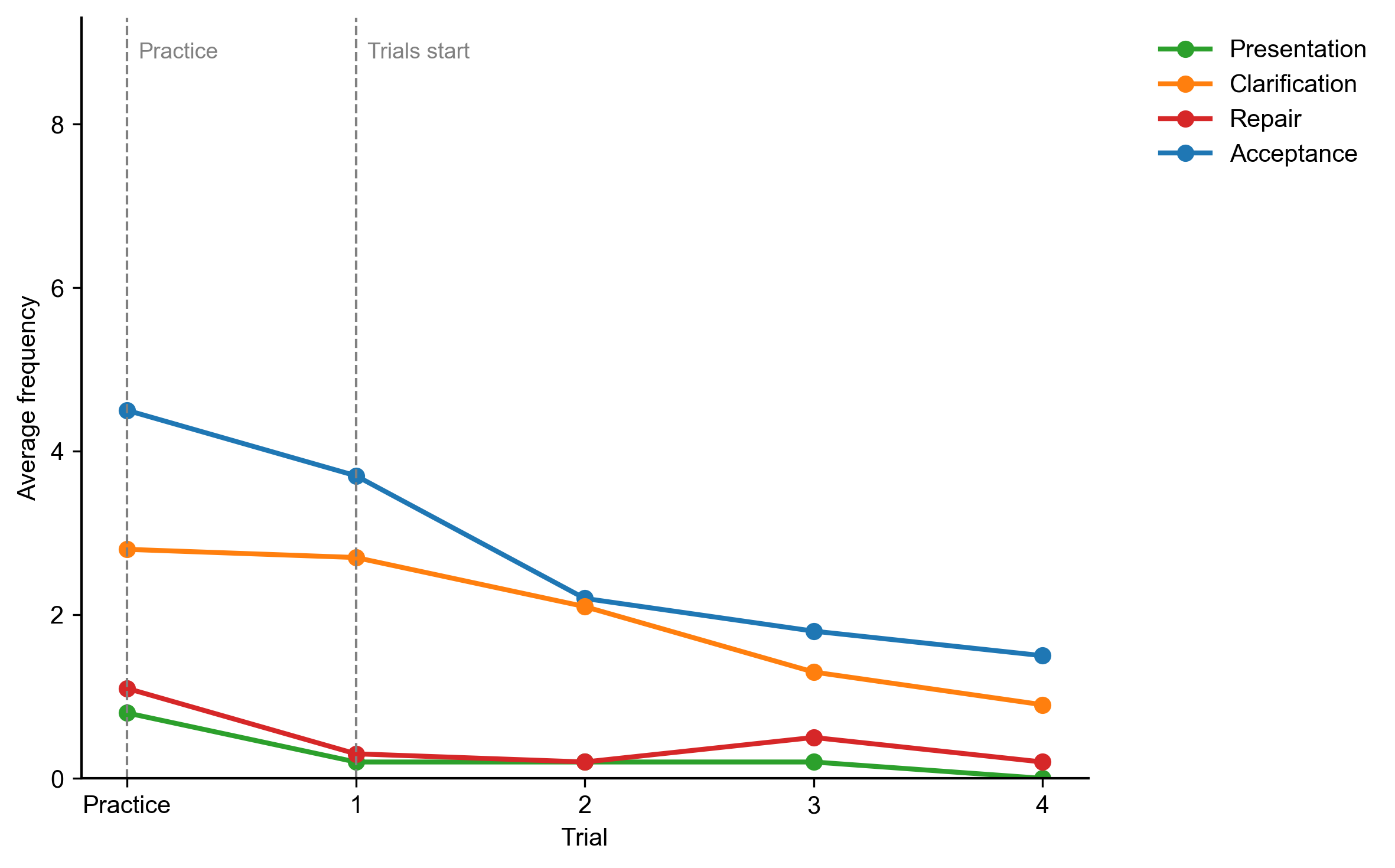}  
    \caption*{(d) Human workers, non-shared view condition.}  
\end{minipage}  
  
\caption{Usage of grounding acts (presentations, clarifications, repairs, and acceptances) in the messages from the human helper and AI worker over the trials in both condition: shared and non-shared view.}  
\label{fig:grounding_types_AI_helper_shared_vs_nonshared_2x2}
\end{figure} 


Fig.~\ref{fig:grounding_types_human_helper_shared_vs_nonshared_2x2} shows the use of the dialogue acts across trials and conditions for human Helpers and AI Workers. We observe a higher use of presentations (more than the four references needed to solve the puzzle) in the earlier trials. In the shared view condition, clarification requests from the AI Worker decrease across trials  (F(1, 28) = 27.63 p $<$ 0.0001), and this is complemented by a (non-significant) peak for repairs at the first puzzle for the human Helper. When the view is not shared, and counter to our expectations, clarifications and repairs are significantly lower than when the Worker has visibility of the Helper's actions. This is the case for both the AI worker (U = 612.5, p $<$ 0.01) and participant Helper (U = 433.0, p $<$ 0.0001). Instead, acceptances dominate the interaction. 

In the shared view, (see Fig.~\ref{fig:grounding_types_AI_helper_shared_vs_nonshared_2x2}) the AI helper frequently uses presentations and acceptances across the trials. Presentations initially decline but, in contrast to the pattern shown for the the human Helper, then show a trend to increase. Repairs show a similar frequency to that for human Helpers, but also show a trend to increase after an initial decrease. We also observe that, when the view is shared, the human Worker initially asks clarification questions but does so less frequently over time (F(1, 28) = 9.4 p $<$ 0.01), while showing a trend for increasing acceptances.

The reduction of clarifications by the Worker over trials seems to align with work by \citet{CLARK19861} and our expectation that mutual understanding is being built. However, this is called into question by the continued high level of repairs by the Helper. This asymmetry suggests that the Helper relies on visual feedback to identify errors, rather than negotiating meaning through messages. This, in turn, aligns with the observations by \citet{Gergle01012013}, that situation awareness supports grounding.

In contrast to our expectations, we do not observe an increase in conversational grounding through clarifications and repairs to compensate for the lack of visual feedback when the view is not shared. Instead, human participants and AI reduce their grounding effort. From the chat message logs, we find that this may originate from the AI often (i) failing to provide clarifications or repairs - even after a clarification request (see Fig.~\ref{fig:example_cycle_3}), and (ii) failing to update its assumptions after repairs (see Fig.~\ref{fig:example_cycle_4}). These issues have been previously observed; \citet{shaikh-etal-2024-grounding} report a lack of grounding initiative in LLMs and \citet{choudhury2025bed} report that LLMs can struggle to maintain and update conversational state across extended interactions.

\begin{figure}[h!]  
\setlength{\tabcolsep}{6pt}    
\begin{tabular}{>{\bfseries}l p{10cm}}  
\rowcolor{gray!20}  
AI & Next, place the beige piece with the criss-cross diamond pattern at the top right corner, next to the pink spiral piece. \\
Human & is that right?  \\
\rowcolor{gray!20}  
AI & I can't see your current progress, but let me guide you to the next step. Place the light pink piece with wavy horizontal lines directly below the pink spiral piece in the left column. \\
\end{tabular}  
\caption{Example of AI helper failing to provide clarification in \textbf{non-shared view} condition.}
\label{fig:example_cycle_3}
\end{figure} 

\begin{figure}[h!]  
\setlength{\tabcolsep}{6pt}    
\begin{tabular}{>{\bfseries}l p{10cm}}  
Human & add the cream piece with horizontal lines to the right of the spiral piece  \\
\rowcolor{gray!20}  
AI & I've placed the cream piece with horizontal lines (ID 18) to the right of the spiral piece at position (2, 0). What should I do next? \\
Human & rotate this piece 90 degrees  \\
$\cdots$ & $\cdots$ \\
Next puzzle &  \\
$\cdots$ & $\cdots$ \\
Human & add the cream piece with horizontal lines in the top left  \\
\rowcolor{gray!20}  
AI & I've placed the cream piece with horizontal lines (ID 18) in the top-left corner at position (0, 0). What should I do next?  \\
Human & rotate this piece 90 degrees  \\
\end{tabular}  
\caption{Example of AI worker "forgetting" previous repairs in \textbf{shared view} condition.}
\label{fig:example_cycle_4}
\end{figure} 

Overall, these findings suggest that human participants seek to address ambiguity and correct mistakes, whereas the AI model is more accepting of what is presented to them. When there is no evidence to highlight that errors are present, in the form of a shared view, the pair eventually work less to mitigate errors through grounding, with clarifications and repairs being less prevalent when there is no shared view, and decreasing over time.

\section{Summary}
\begin{table}[htbp]  
\centering  
\setlength{\tabcolsep}{6pt}  
\begin{tabular}{p{6.5cm} p{6.5cm}}  
\toprule  
\textbf{Confirming our expectations} & \textbf{Not confirming our expectations} \\  
\midrule  
\textbf{Task-level performance:} Higher puzzle success in the shared view condition compared to the non-shared view condition, consistent with prior work showing the benefits of shared visual context.   
&  
\textbf{Task-level performance:} No learning effect over trials in either condition; puzzle accuracy does not improve with experience, contrary to expectations of increased coordination over time.  
\\[6pt]  
  
\textbf{Communication effort (shared view):} Human Helpers reduce communication effort over trials in the shared view condition, indicating increased efficiency and alignment.  
&  
\textbf{Communication effort (non-shared view):} Overall communication effort does not increase in the non-shared view condition, contradicting prior findings that predict higher communicative effort when visual context is unavailable.  
\\[6pt]  
  
\textbf{Referential convergence:} Human participants adapt to identifier-based references introduced by the AI Worker, indicating asymmetric convergence toward AI-proposed referential conventions.  
&  
\textbf{Lexical convergence:} Jointly used vocabulary remains small and decreases over trials, providing limited evidence for the emergence of a shared lexicon.  
\\[6pt]  
  
\textbf{Definite reference use:} Both human participants and AI predominantly use definite references from the beginning, consistent with an assumption of mutual intelligibility supported by visual feedback.  
&  
\textbf{Least collaborative effort:} Identifier use does not yield symmetric efficiency gains, as the AI continues to combine identifiers with descriptive references.  
\\[6pt]  
  
\textbf{Conversational grounding (shared view):} Clarification requests by Workers decrease over trials, aligning with grounding theory predictions that explicit negotiation declines as mutual understanding accumulates.  
&  
\textbf{Conversational grounding (non-shared view):} Clarifications and repairs do not increase to compensate for the lack of shared view; instead, grounding activity decreases over time.  
\\[6pt]  
  
\textbf{Situation awareness:} Helpers rely on visual feedback to detect and repair errors directly, consistent with prior findings that situation awareness supports grounding.  
&  
\textbf{AI grounding behavior:} The AI frequently fails to initiate or respond appropriately to clarifications and does not reliably update assumptions after repairs, limiting grounding progress.  
\\  
\bottomrule  
\end{tabular}  
\caption{Summary of the findings from the confirmatory study in relation to expected observations across task-, object-, and communication-level analyses for common ground.}  
\label{tab:expectations_summary}  
\end{table}  

Tab.~\ref{tab:expectations_summary} summarises the results from the previous section and groups them into two categories: (i) findings that confirm expectations derived from prior human communication studies on common ground, and (ii) findings that do not confirm these expectations, instead diverging from observed human communication patterns.

\section{Conclusion}
We introduced a novel benchmark for evaluating common ground in human-AI interaction. Building on classical paradigms for studying common ground human communication, we designed a collaborative puzzle matching task which involved messaging and visual feedback. The benchmark provides a structured and reusable setup to study mutual understanding, situation awareness, and grounding processes in human-AI interaction, allowing research into the development of common ground between people and AI.

In a confirmatory study, we demonstrated how the benchmark operationalises concepts from common ground theory, enabling measurements of learning effects and communicative efficiency, through coordination using referential conventions, conversational grounding, and situation awareness. Using a specific fixed LLM, GPT4.1, interacting with human participants, we illustrated how patterns of human-AI interaction align with or differ from those that illustrate grounding in human communication. To avoid overly strong or general conclusions, we emphasise that these findings are contingent to the particular AI model, i.e., GPT4.1. Other prompting strategies, different interfaces, or alternative current or future models may exhibit different behaviours and result in different patterns.
 
Emphasising that our findings are specific to GPT4.1, the study results reveal both human and AI specific interaction patterns. GPT4.1 uses visual feedback to repair misunderstandings. When there is no shared view, it sometimes ignores or does not build upon clarification requests, and accepts statements without attempting verification. It struggles to maintain the state of the puzzle task, with the result that repairs continue to occur across trials. It does not adapt to participants' vocabulary; most noun phrases are used solely by GPT4.1 and take the form of detailed descriptive references to the puzzle pieces. Still, collaboration can be successful if situation awareness is supported.

People, too, rely on visual feedback to repair misunderstandings. However, in contrast to the model, they try to engage, initially, with clarification requests. These decrease over time; we speculate that this is because the model does not build on them, and there is no sense that they become part of the the common ground. The result of this is that people too repair less and accept more over time, which impacts performance. People do adapt to AI‑initiated reference forms, such as puzzle piece identifiers, even though this likely incurs cognitive load. 

The study confirms several of our expectations from theories and studies of human communication. The results that contradict the expected interaction patterns can be explained by model specific issues reported in previous works e.g., limited grounding efforts or ineffective maintenance of belief states, and may also reflect the use of text as the communication medium. Further studies could study human-AI interaction via spoken language, or with anthropomorphised interfaces, to explore how this impacts people's communication patterns and willingness to perform grounding acts. By introducing our new benchmark and considering the study results, while specific to GPT4.1, we hope to establish a new way to evaluate the development of common ground in human-AI interaction.

\bibliographystyle{apalike}
\bibliography{bib}





\end{document}